\documentclass[]{aa}  
\usepackage{graphicx}
\graphicspath{{../../Plots_nPoly/}}
\usepackage{txfonts}
\usepackage{natbib,twoopt}
\usepackage[breaklinks=true]{hyperref} %% to avoid \citeads line fills
\bibpunct{(}{)}{;}{a}{}{,}             %% natbib format for A&A and ApJ
\newcommand{\kms} {km\,s$^{-1}$}

\newcommand{\Msun} {$\mbox{M}_{\sun}$}

\newcommand{\kpc}{\,{\rm kpc}}
\newcommand{\mpc}{\,{\rm Mpc}}
\newcommand{\Mperyr} {$\mbox{M}_{\sun}\,$yr$^{-1}$}

\begin{document} 

\title{Exploring the mass assembly of the early-type disc galaxy NGC\,3115 with MUSE}

\author{A.\,Gu\'{e}rou
          \inst{1,2,3,*}
          \and
          E.\,Emsellem
          \inst{3,4}
          \and
          D.\,Krajnovi\'{c}
          \inst{5}
          \and
          R.\,M.\,McDermid
          \inst{6,7}
          \and
          T.\,Contini
          \inst{1,2}
          \and          
          P.\,M.\,Weilbacher
          \inst{5}
          }

\institute{
			IRAP, Institut de Recherche en Astrophysique et Plan\'{e}tologie, CNRS, 14, avenue Edouard Belin, F-31400 Toulouse, France
		\and
			Universit\'{e} de Toulouse, UPS-OMP, Toulouse, France
		\and
			European Southern Observatory, Karl-Schwarzschild-Str. 2, D-85748 Garching, Germany
			\email{aguerou@eso.org}
		\and
			Universite Lyon 1, Observatoire de Lyon, Centre de Recherche Astrophysique de Lyon and Ecole Normale Sup\'{e}rieure de Lyon, 9 avenue Charles Andr\'{e}, F-69230 Saint-Genis Laval, France
		\and
			Leibniz-Institut f\"{u}r Astrophysik Potsdam (AIP), An der Sternwarte 16, D-14482 Potsdam, Germany
		\and
			Department of Physics and Astronomy, Macquarie University, Sydney NSW 2109, Australia
		\and
			Australian Astronomical Observatory, PO Box 915, Sydney NSW 1670, Australia
             }

   \date{Received ... ; accepted ...}

%%%%%%%%%%%%%%%%%%%%%%%%%%%%%%%%%%%%%%%%%%%%
% \abstract{}{}{}{}{} 
% 5 {} token are mandatory
  \abstract
  {We present MUSE integral field spectroscopic data of the S0 galaxy NGC\,3115 obtained during the instrument commissioning at the ESO Very Large Telescope (VLT). We analyse the galaxy stellar kinematics and stellar populations and present two dimensional maps of their associated quantities. We thus illustrate the capacity of MUSE to map extra-galactic sources to large radii in an efficient manner, i.e.\,, $\sim$4 $R_e$, and provide relevant constraints on its mass assembly. We probe the well known set of substructures of NGC\,3115 (its nuclear disc, stellar rings, outer kpc-scale stellar disc and spheroid) and show their individual associated signatures in the MUSE stellar kinematics and stellar populations maps. In particular, we confirm that NGC\,3115 has a thin fast rotating stellar disc embedded in a fast rotating spheroid, and that these two structures show clear differences in their stellar age and metallicity properties. We emphasise an observed correlation between the radial stellar velocity, $V$, and the Gauss-Hermite moment, $h_3$, creating a ``butterfly'' shape in the central 15\arcsec\ of the $h_3$ map. We further detect the previously reported weak spiral and ring-like structures, and find evidence that these features can be associated with regions of younger mean stellar ages. We provide tentative evidence for the presence of a bar, despite the fact that the $V$-$h_3$ correlation can be reproduced by a simple axisymmetric dynamical model. Finally, we present a reconstruction of the two dimensional star formation history of NGC\,3115 and find that most of its current stellar mass was formed at early epochs (>12~Gyr ago), while star formation continued in the outer (kpc-scale) stellar disc until recently. Since $\textit{z}\sim$2, and within $\sim$4 $R_e$, we suggest that NGC\,3115 has been mainly shaped by secular processes.}
  
  
  
   \maketitle
%
%%%%%%%%%%%%%%%%%%%%%%%%%%%%%%%%%%%%%%%%%%%%
\section{Introduction}

Key mechanisms related to galaxy formation and evolution, such as stellar feedback, merging, gas and satellites accretion, leave their footprints on galaxy morphology, dynamical structures and stellar components. Both the cores and the outskirts of galaxies thus contain crucial information that should help us understanding the onset and assembly of galactic systems. Simulations have shown that the above-mentioned physical processes create dramatic changes in a galaxy's stellar kinematics \citep[e.g.\,,][]{Hoffman2010, Bois2011} and stellar population \citep{Hopkins2009} between their inner and outer regions.

One of the strong predictions from modern galaxy formation scenarios is that the relative contributions and timing of in-situ and ex-situ star formation should appear as a relative change in the stellar content and structures from the centre to the outer parts of galaxies \citep{Dekel2009, Zolotov2009, Naab2009, Oser2010, Font2011, Lackner2012, Navarro-Gonzalez2013, Rodriguez-Gomez2015}. The characteristic radius at which this transition is expected to happen and be detected is still debated. Simulations of gas-rich mergers~\citep{Hoffman2010} show that kinematical transitions should occur between 1--3 galactic effective radius ($R_e$), whereas cosmological simulations~\citep{Hirschmann2015} including, e.g.\,, stellar winds and gas/satellites accretion found that the transition radius could be pushed outwards between 4--8~$R_e$ (based on the metallicity gradient). Photometric studies have started to tackle this issue outside the local group~\citep{Martinez-Delgado2010, Forbes2011, VanDokkum2014} and recently large surveys \citep{Roediger2011, Duc2015} have taken a detailed quantitative look at early-type galaxies revealing prominent stellar streams, tidal-tails, stellar shells, etc, that stress again the importance of merging, gas and satellites accretion in the assembly histories of galaxies. Photometric studies are powerful tools to study the relics of mass assembly in the low surface brightness outer regions of galaxies, where long dynamical timescales can preserve such signatures up to several gigayears. Spectroscopic studies, however, allow precise kinematic and stellar population signatures to be determined, with the potential to reveal how and when the stellar mass formed over cosmological timescales.

Milestones have been reached via Integral Field Unit (IFU) surveys such as ATLAS$^{\rm 3D}$ \citep{Cappellari2011a}, suggesting a new classification of early-type galaxies based on their dynamical status, or CALIFA \citep{Sanchez2012,Garcia-Benito2015} which systematically covered a diameter selected sample of nearby galaxies. These surveys provide spatial resolutions often reaching down to 100~pc or lower, allowing the study of star formation regions, inner stellar and gaseous structures such as circumnuclear discs, or decoupled cores. They, however, probe typically radii up to one or two effective radii only. Other successful surveys like SLUGGS \citep{Brodie2012,Brodie2014} were designed to probe galaxies to much larger radii. These are time consuming campaigns and often miss either spatial resolution and/or sample size. Challenging surveys like SAMI \citep{Croom2012,Bryant2015}, or MaNGA \citep{Bundy2015} are trying to fill that gap with impressive samples of thousands of galaxies, but unfortunately will lack the combined radial coverage and spatial resolution to resolve sub-kiloparsec structures and large scale properties simultaneously.

The new IFU, MUSE \citep{Bacon2010}, mounted on the UT4 of the Very large Telescope (VLT), has been designed to offer a relatively large field of view (1\arcmin~$\times$~1\arcmin\,), with good spatial sampling (0\farcs2) and very high throughput, making it a powerful tool for mapping nearby extra-galactic targets to large radii in a reasonable amount of time. Its capacity to perform large mosaics \citep{Weilbacher2015} as well as its extreme stability, calibration quality and high sensitivity~\citep{Bacon2015} has already been demonstrated and makes this instrument one of the best existing IFU instrument to study detailed stellar and ionised gas properties (kinematics and population) of extra-galactic targets.

NGC\,3115 was observed during the first commissioning run of the MUSE spectrograph, with the intention of demonstrating the instrument's capabilities and testing its performance. Thanks to such commissioning data, the strategy of observations, calibration scheme, and instrument setup were developed further, in parallel of an updated data reduction pipeline, improving dramatically the output data quality of the MUSE instrument \citep{Weilbacher2015_proc}. NGC\,3115 is the closest S0 galaxy to the Milky Way ($d=9.8\mpc\,$,~\citealt{Cantiello2014}), with a subsequently large apparent diameter ($\mu{_B}$\,=\,25\,mag.arcsec$^{-2}$~$\sim$~8\arcmin). It is bright and almost edge-on~($i=86^{\circ}$,~\citealt{Capaccioli1987}) with very little dust and gas~\citep{Li2011, Li2013}. It has a smooth surface brightness distribution, making it ideal to test the ``spectro-imager'' and mosaicing abilities of MUSE. NGC\,3115 is also a very interesting object with a complex set of substructures, including a nuclear disc and an outer (kpc-scale) disc~\citep{Capaccioli1987, Nieto1991, Scorza1995, Lauer1995, Emsellem1999}, possible rings and spirals~\citep{Norris2006, Michard2007, Savorgnan2015}, a rapidly rotating spheroid \citep{Arnold2011, Arnold2014}, a large globular clusters system exhibiting a clear bi-modality \citep{Kuntschner2002, Brodie2012, Jennings2014a, Cantiello2014}, a central super-massive black hole~\citep{Kormendy1992, Kormendy96, Emsellem1999} and a few X-ray sources \citep{Wong2011, Wrobel2012}.

This paper  is organised as follow. In \S~\ref{sec:ObsDataRed} we present the observations and our data reduction, especially our sky subtraction method. In \S~\ref{sec:kinAna} we explain our stellar kinematics extraction and analysis, and in \S~\ref{sec:pop}, the stellar population and star formation history that we derive. In \S~\ref{sec:discussion} we discuss the different substructures of NGC\,3115, including the possible existence of a bar, that we use to constrain the formation and evolution scenarios of the galaxy within $\sim$4 $R_e$. We conclude in \S~\ref{sec:conclusion}.

%%%%%%%%%%%%%%%%%%%%%%%%%%%%%%%%%%%%%%%%%%%%
\section{The MUSE NGC\,3115 dataset}
\label{sec:ObsDataRed}

\subsection{Observations and data reduction}
\label{subsec:obs}
We present here the first IFU observations, obtained with MUSE, of the early-type S0 galaxy NGC\,3115. These observations were obtained during the first instrument commissioning (ESO program 60.A-9100(A)). Before describing the data reduction process itself, it is important to emphasise that the observations were obtained in conditions far from optimal: in terms of the definition of the observing blocks, the strategy for the sky exposures, the calibration strategy, the sky background (moon illumination), etc. As a consequence, and despite a careful data reduction, strong systematics still persist and may bias the analysis. The impact of systematic errors in the data resulting from these non-optimal aspects of the observations are considered in detail in section \S\,\ref{subsec:skysub}, \S\,\ref{subsec:kinSystematics} and \S\,\ref{subsec:popMeasure}.

\begin{figure}
   \centering
   \includegraphics[width=\columnwidth]{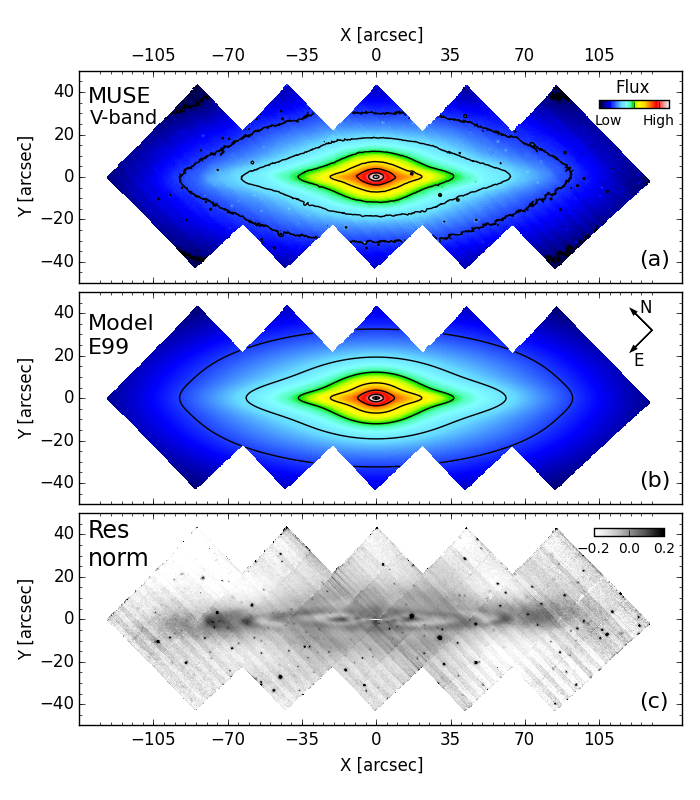}
   \caption{Panel~(a):~MUSE reconstructed V-band image of NGC\,3115 obtained through the MUSE pipeline.  Panel~(b):~Photometric V-band model of NGC\,3115 from \cite{Emsellem1999}. The model flux has been scaled to the MUSE V-band image, which is in arbitrary unit. Panel~(c):~Normalised residuals of the MUSE reconstructed V-band image, obtained by subtracting the \cite{Emsellem1999}\,~model. The overall flatness of the residuals shows the good quality of our data reduction process (see \S\,\ref{subsec:skysub}). The significant residual structures~(inner-disc, spirals, etc.\,) are discussed in \S\,\ref{subsec:substructures}.}
   \label{Fig:phot_res}
    \end{figure}

NGC\,3115 was observed during the night of the 8$^{th}$ of February 2014 using MUSE nominal mode (WFM-N) that allows a continuous wavelength coverage from 4750~--~9300\,\AA\,~with a varying resolution of R$=$2000~--~4000. These data were designed as a first test of the mosaicing abilities of MUSE using an extended (bright) target which showed substructures and had ample published imaging and spectroscopic data to compare with. Five exposures of only 10 minutes each were obtained, each exposure overlapping with its neighbors over a quarter of the field of view~(i.e.\,,~30\arcsec\ $\times$~30\arcsec), with the central exposure being centered on the galaxy nucleus. The data obtained cover $\sim$4\,R$_e$ (R$_e$\,$\sim$35\arcsec\,, \citealt{Arnold2014}) along the NGC\,3115 major-axis (see Fig.~\ref{Fig:phot_res}). One extra exposure of 10 minutes was taken off-target in the middle of the observing block, 4\arcmin\,~away from the target toward the east, to be able to estimate the sky level contribution. Neither dithering nor rotations were done between exposures. The data were taken under good weather conditions with a recorded DIMM seeing (Full Width at Half Maximum, FWHM) varying between 0\farcs7 and 0\farcs8.\

We used the MUSE reduction pipeline (v1.0, \cite{Weilbacher2012pipeline}, Weilbacher~et~al.~in~prep.) to reduce the data and used the associated standard calibration files (bad pixels, filters list, lines and sky lines catalogue, extinction table, vignetting), except for the ``astrometry'' and ``geometry'' solutions, that come from the first commissioning run (since the IFUs were re-aligned since then). All exposures (on- and off-target) were reduced in the same way, as follows: bias subtracted, flat-fielding and illumination corrected (using a twilight exposure), wavelength calibrated, using master calibration files built from calibration files which are the closest in time to the target exposures. We checked that the temperature did not vary significantly (less than one or two degrees) in between calibration files and on- and off-target exposures. One standard star was observed at the beginning of the night, GD\,71, that we used to flux calibrate the science exposures.\

%_________________________
% Figure Photometry

\subsection{Sky subtraction}
\label{subsec:skysub}

The observations were taken with 81\% of moon illumination (67 degrees away) during the moon set, leading to a very high sky background level with very strong variations between each on-target exposure. As mentioned above, only one sky exposure was obtained and thus available for the five on-target exposures. To subtract the correct sky level background to each on-target exposure, we developed the following method. 

We first sky subtracted each science exposure individually using the direct MUSE pipeline measurement of the sky spectrum from the sky exposure. We then used the overlapping area between each on-target exposure to estimate the relative spectral variation of the sky level from one science exposure to another. We used four areas of 25$\times$25 spaxels (i.e.\,,~$\sim$5\arcsec$\times$~5\arcsec) on each exposure for such a comparison. We obviously made sure, beforehand, that all exposures were aligned using foreground stars and/or globular clusters. The time variation of the sky lines being quite complex, we only estimated the lunar continuum variation by fitting a third order polynomial to the spectral variation of the continuum. We then scaled that modeled difference accordingly, to create a total spectrum of the sky continuum variation (i.e.\,,~smoothing the noise added by the subtraction process) between two overlapping on-target exposures. The sky lines are not corrected through this process (i.e.\,,~not taken into account by the scaling of the sky level), and we will thus mask these for further analysis of the spectra.

Since the sky exposure and on-target exposures have the same exposure time, we first applied this method to the two exposures framing the sky exposure itself and added (or subtracted) half of the measured sky continuum variation to the initial estimation. We then performed the sky subtraction of these two exposures via the MUSE pipeline, using the new respective estimation of the sky continuum and letting the pipeline fitting the sky emission lines using the available sky lines list. We then propagated the method described above using the reduced corrected on-target exposures. To validate our sky subtraction method, we compared the obtained spectra in the area used to estimate the sky level variation between two overlapping exposures. There, we measured a mean spectral variation (for the same physical area on-target) of the order of three to five percent. 

We performed the merging of the five exposures through the MUSE pipeline and obtained a fully-reduced cube, with its associated noise as estimated and propagated through the MUSE pipeline. To confirm once again the good performance of our method, we subtracted a photometric V-band model of NGC\,3115~\citep[published in][]{Emsellem1999} from the MUSE reconstructed V-band image~(see Fig.~\ref{Fig:phot_res}) and observed flat normalised residuals over the full coverage of the MUSE pointings, i.e.\,,~$\sim5$\% absolute fluctuation on average. The significant intrinsic residual structures visible on panel \textit{(c)} of Fig.~\ref{Fig:phot_res} (inner disc, spirals, rings, etc.\,) are discussed in \S\,\ref{subsec:substructures}.

\subsection{Voronoi binning of the MUSE cube}
\label{subsec:binning}
Prior to any analysis of the data, we spatially binned our reduced MUSE cube to increase and homogeneise the signal-to-noise ratio (S/N) of the spectra. We used the adaptive spatial binning software developed by \cite{Cappellari2003Bin}\footnote{\label{note:notecapcode}Python codes publicly available at the following address: http://www-astro.physics.ox.ac.uk/~mxc/software/}, based on Voronoi tessellation. We estimated the original S/N of each individual spectrum by using the variance spectra produced by the MUSE pipeline. We used a median within a relatively narrow spectral window, between 5450\,\AA\,~and 5550\,\AA\, which is clean of strong sky emission lines, as well as representative of the wavelength range we will use for extracting the stellar kinematics and population properties.

We chose a target S/N of 50 per \AA\,~for each spatial bin (spaxels), which offers good data quality to perform a robust stellar kinematics and population extraction (S/N of 40 is usually enough, \citealt{Cappellari2011a}), while keeping reasonable spatial sampling. The adaptive spatial binning delivered more than 60\% of the bins with a S/N equal or greater than 50, a general scatter of $\sim$7 and less than 0.1\% of the bins with a S/N lower than 40. The central $\sim$40\arcsec$\times$10\arcsec\ having originally a S/N greater than 50, were kept unbinned.

%%%%%%%%%%%%%%%%%%%%%%%%%%%%%%%%%%%%%%%%%%%%
\section{Stellar kinematics analysis}
\label{sec:kinAna}
%__________________________________________

%_________________________
% Figure kin: fit ppxf
\begin{figure*}
   \centering
   \includegraphics[width=\textwidth]{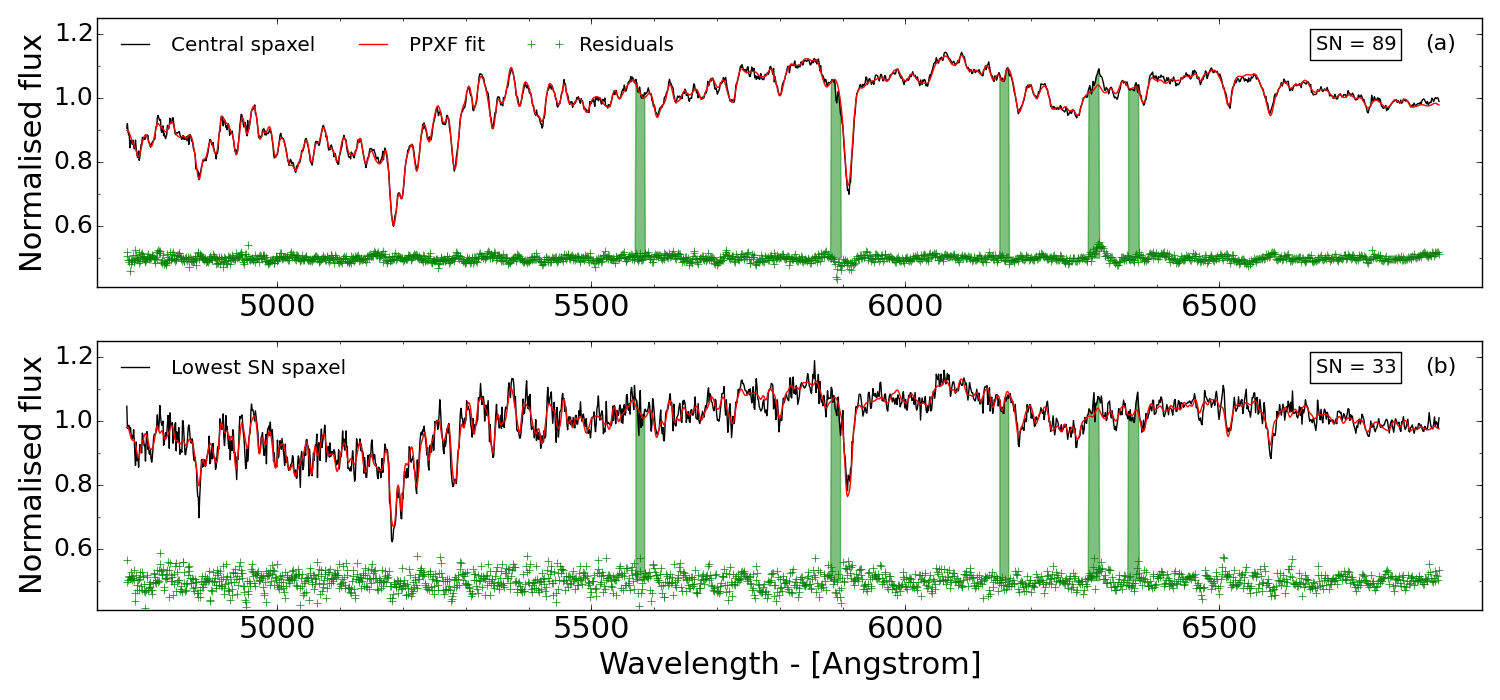}
   \caption{Stellar kinematics fits: (a) of the central spaxel, and (b) the lowest S/N spectrum of our MUSE NGC\,3115 data cube. The fits were performed with the pPXF software \citep{Cappellari2004ppxf} using the MILES stellar library~\citep{SanchezBlazquez2006MILES,Falcon2011}, between 4760~--~6850\,\AA\,. The black lines show the MUSE spectra, the red lines the best fits, the shaded green areas are the wavelength ranges not fitted (i.e.\,,~corresponding to the sky lines regions), and the green crosses show the fit residuals (shifted upwards by 0.5 along the y axis). The upper-right box in each panel indicates the respective S/N of the MUSE spectrum showed.}
   \label{Fig:fit_kin}
    \end{figure*}

%_________________
% Figure kin: maps
\begin{figure*}
   \centering
   \includegraphics[width=\textwidth]{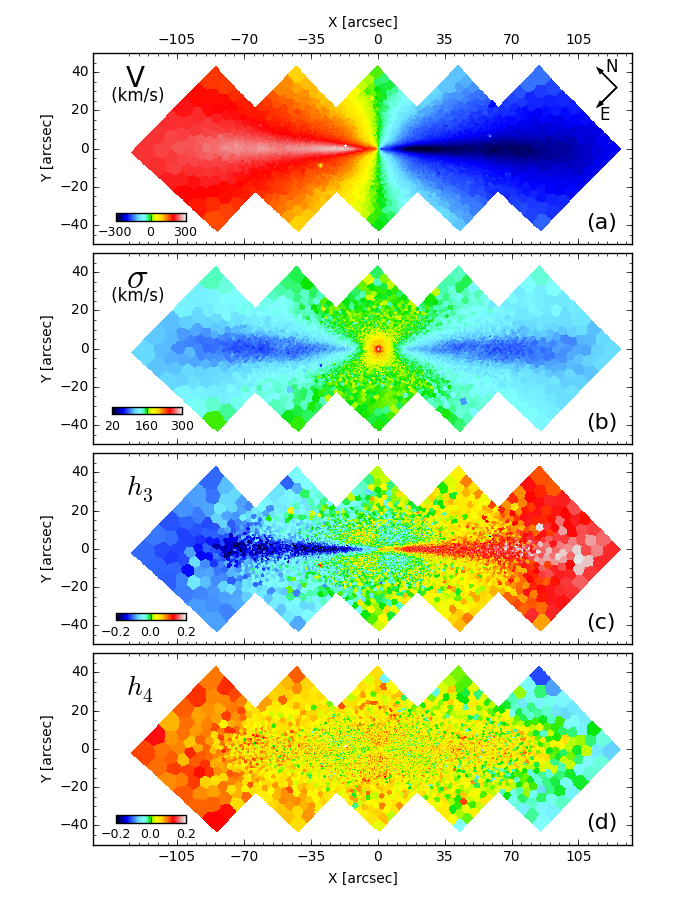}
   \caption{Maps of the first four Line of Sight Velocity Distribution (LOSVD) moments: (a) mean stellar radial velocity, V ; (b) mean stellar velocity dispersion, $\sigma$ ; (c) third order Gauss-Hermite moment, $h_3$ ; (d) fourth order Gauss-Hermite moment, $h_4$. The color scheme for each panel is indicated on its bottom left corner. The kinematics maps clearly show a thin, kinematically cold, fast-rotating disc, embedded in a kinematically hotter, rotating spheroid. Note the correlation between $h_3$ and $V$ in the central 15\arcsec\,,~away from the major-axis, suggesting a central bar that we discuss in \S\,~\ref{subsec:h3}.}
   \label{Fig:kin}
    \end{figure*}

%__________________________________________
\subsection{Stellar kinematics extraction and estimation of the systematic errors}
\label{subsec:kinSystematics}
To measure the stellar kinematics of NGC\,3115 from the MUSE cube, we used the penalised pixel-fitting method (pPXF)$^{\ref{note:notecapcode}}$~developed by \cite{Cappellari2004ppxf}. pPXF uses a stellar library to build an optimal template that will best fit the observed spectrum, using Gauss-Hermite functions. We used the full MILES\footnote{MILES stellar library can be downloaded here: http://miles.iac.es/pages/stellar-libraries.php} stellar library \citep{SanchezBlazquez2006MILES, Falcon2011} covering the wavelength range 3525~--~7500\,\AA\ with a constant spectral resolution of 2.50\,\AA\,FWHM, similar to but lower than the MUSE spectral resolution. Since the difference in spectral resolution between the MUSE data and the MILES stellar library (0.2\,\AA\,FWHM at 5000\,\AA\,,~i.e.\,, $\sigma$=5~\kms\,) is significantly lower than NGC\,3115 velocity dispersion ($\sigma$>70~\kms\,), we did not convolve the stellar library to the same spectral resolution as the observed spectra.

We derived the first four order moments of the line of sight velocity distribution (LOSVD), namely V, $\sigma$, $h_3$ and $h_4$, for each spaxel of the MUSE cube, in the following manner. We first selected a collection of templates among the MILES library, from the fitting results of a representative MUSE spectrum of NGC\,3115. Such a spectrum was obtained by stacking the spectra within a circular aperture of 5\arcsec\ (in radius), centred on the galaxy photometric centre. Then, we used this combination of templates to fit each spaxel of the MUSE data. We set up pPXF to use additive polynomials of the 8$^{th}$ order and the default value of penalisation (0.4). We restricted the wavelength range fitted to 4760~--~6850\,\AA\,,~to avoid spectral regions highly contaminated by sky lines ($\lambda\,$>7000\,\AA) and masked five strong sky emission lines at 5577.4\,\AA\,~;~5889.9\,\AA\,~;~6157.5\,\AA\,~;~6300.3\,\AA\,~and 6363.7\,\AA\,(see \S\,\ref{subsec:skysub}). Given the lack of evidence of any dust or gas in NGC\,3115, either from the literature or from our MUSE spectra, we did not mask for any potential ionised gas emission lines. Using equation (1) of \cite{Sarzi2006}, we estimate our gas emission lines detection limit to an equivalent width of 0.26\,\AA\, for H$\beta$ and [OIII]. For illustration, two examples of the performed fits are presented in Fig.~\ref{Fig:fit_kin}, including the central spaxel (with a S/N$\sim$90) and the lowest SN spaxel (S/N$\sim$33) of the spatially binned MUSE observations.
 
To quantify the impact of our fitting parameters on our results, we performed similar analyses but varying one parameter at a time. Firstly, we used additive polynomials of the 4${th}$ order, commonly used for fitting spectra on narrower, SAURON-like, wavelength range. Secondly, we masked all the potential emission lines from the galaxy, e.g., H$\beta$, [OIII], H$\alpha$, [NII] and [SII]. And thirdly, we fitted the MUSE spectra on a restricted wavelength range (5000~--~5500\,\AA) that contains the Fe lines and the Mg$b$ triplet, which strongly constrain on their own the derived stellar kinematics. In all cases, we measured similar kinematics structures, with mean differences of only a few kilometers per second for the two first LOSVD moments, V and $\sigma$, and mean differences of 0.001 to 0.02 for $h_3$ and $h_4$.

We should, however, emphasise that there are still clear systematic errors regarding, mostly, the third and fourth order moments $h_3$ and $h_4$. The corresponding maps show some odd features, as for example a clear offset of the $h_4$ level at the edges of the field (see Fig.~\ref{Fig:kin} and Fig.~\ref{Fig:kin_cut}). This is not surprising considering the original strong variation of the sky background (see \S\,\ref{subsec:skysub}), both between exposures and also within the field of view of individual exposures (residual gradient). Assuming symmetric (odd or even) moment maps for $h_3$ and $h_4$ (with respect to the photometric minor-axis), we estimate these systematic errors to 0.02 for $h_3$ and 0.05 for $h_4$.

\subsection{Stellar kinematics results}
\label{subsec:kin}

%_________________
% Figure kin: cut kin
\begin{figure}
   \centering
   \includegraphics[width=\columnwidth]{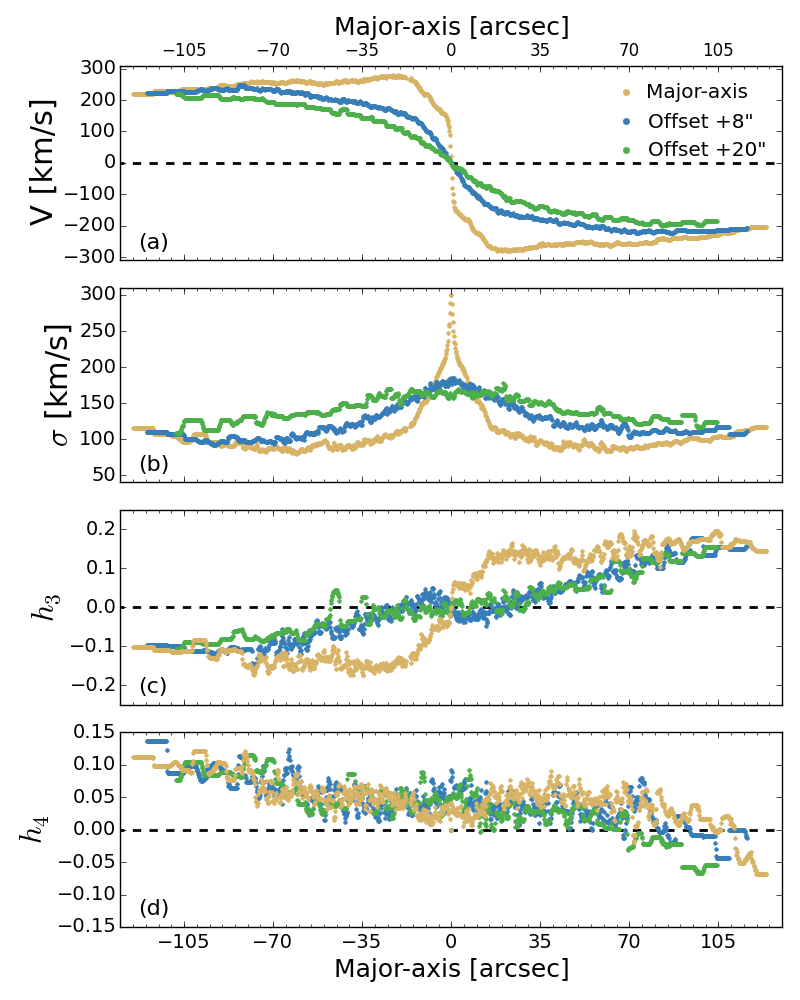}
   \caption{One dimensional curve of the first four Line of Sight Velocity Distribution (LOSVD) moments: (a) Stellar radial velocity, V ; (b) Stellar velocity dispersion, $\sigma$ ; (c) third order Gauss-Hermite moment, $h_3$ ; (d) fourth order Gauss-Hermite moment, $h_4$. The points have been extracted from the two dimensional maps of the respective quantities (see Fig.~\ref{Fig:kin}), by taking the light-weighted mean value within a slit of 2\arcsec\ wide, aligned with the major-axis (orange points), and offset by +8\arcsec\ (green points) and +20\arcsec\ (blue points) along the minor-axis.}
   \label{Fig:kin_cut}
    \end{figure}

%__________________________________________

We present in Fig.~\ref{Fig:kin} the stellar kinematics maps of NGC\,3115, obtained as described in \S\,\ref{subsec:kinSystematics} from our MUSE reduced cube. The mean radial stellar velocity map (Fig.~\ref{Fig:kin}, panel~\textit{a}) exhibits a fast rotating, thin disc-like structure embedded in a slower spheroidal component, but still rotating fairly rapidly. The thin disc-like structure has an apparent vertical extent (with respect to the major-axis) widening from a few arcseconds near the centre, to $20$\arcsec\ (i.e.\,,~$\sim$930~pc) at about two effective radii (i.e.\,,~$\sim$70\arcsec\ or $\sim$3.4\kpc), where it becomes blended with the spheroidal part of the galaxy. The thinness of the kinematic disc also confirms that NGC\,3115 is viewed quasi edge-on. Along the major-axis, the mean radial stellar velocity curve exhibits a double maximum (see Fig.~\ref{Fig:kin_cut}, panel \textit{a}) with a first steep rising up to 150~\kms\ within the central 2\arcsec\,, a first plateau between 2~--~7\arcsec\,,~and a second maximum at $\sim$9\arcsec\ peaking at $\sim$200~\kms\,,~before rising up to the maximum rotation of 280~--~290~\kms\ at $\sim$20\arcsec\ from the center. At larger radii, the mean radial stellar velocity flatten to 205~\kms. Further away from the major-axis, i.e.\,, just above the equatorial plane, the velocity gradient across the minor axis becomes significantly shallower, representative of the rotation curve of the spheroidal component.

This superposition of a disc-like structure embedded in a rotating spheroid is also beautifully revealed in the velocity dispersion map (Fig.~\ref{Fig:kin}, panel~\textit{b}): the lowest velocity dispersion values ($\sim$80~--~90~\kms) match precisely the spatial extent of the disc, and rapidly increase (up to $\sim$150~\kms) moving off the major axis, corresponding to where the spheroidal component dominates. We also notice a transition zone between the dynamically cold disc and spheroid, where the velocity dispersion values stay $\sim$100~--~110~\kms\,. The stars associated with this relatively cold and fast rotating component could be interpreted as a "thick" disc-like structure, although it is difficult to identify this just from the kinematics itself. In the centre of the galaxy, the velocity dispersion peaks at 321~\kms\ (based on individual spaxel) that is very probably partly the sign of a central dark mass~\citep{Kormendy1992, Kormendy96, Emsellem1999}. We also notice that the velocity dispersion within the central 15\arcsec\ is higher than in the disc component, whereas the disc-like structure goes all the way to the centre in the velocity map (Fig.~\ref{Fig:kin_cut}, panel~\textit{a}). It is known from photometry that NGC\,3115 has a nuclear disc~\citep{Lauer1995, Kormendy96} of $\sim$3\arcsec\ in radius and an inner ring~\citep{Savorgnan2015} of $\sim$15\arcsec\ (see also \S\,~\ref{subsec:substructures}). What we see in the $\sigma$ map is therefore the superposition along the line-of-sight (LOS) of four (maybe five) kinematic components: the nuclear-disc, the inner-ring, the thin disc (maybe a thick disc), and the spheroid, that results in an increase of the velocity dispersion (each component having a different rotational speed). At large radii, i.e.\,,~greater than 80\arcsec, slightly larger than two effective radii, all these kinematic structures fade into one single rotating structure, as the $V$ and $\sigma$ maps flatten uniformly. We can therefore constrain the apparent radial extent of the thin stellar disc to $\sim$~80\arcsec\ (i.e.\,,~$\sim$3.8\kpc). This outer structure forms a smooth and continuous extension of the spheroidal component that dominates off the equatorial plane ($\sim$20\arcsec) at all radii.

Fig.~\ref{Fig:kin}~(panel~\textit{c}) shows the $h_3$ map of NGC\,3115, the third Gauss-Hermite moment. $h_3$ represents the skewness of the LOSVD. A positive (negative) $h_3$ value means that the LOSVD skew towards lower (higher) velocities with respect to the mean \citep{VanderMarel1993, Gerhard1993}. Over most of the covered field of view, NGC\,3115 $h_3$ moment is anti-correlated with respect to its stellar radial velocity, $V$. Such a $V$-$h_3$ anti-correlation is commonly observed (see \citealt[][and references therein]{Bender1994, Krajnovic2008}) and expected from fast rotating disc-like structures embedded within a slower rotating, more spheroidal stellar component, as in NGC\,3115. Fig.~\ref{Fig:kin_cut} (panel \textit{c}) shows that $h_3$ exhibits a small turn-over at a radius of $\sim$2\arcsec\ and a plateau at $\sim$15\arcsec\ along the major-axis, that coincide perfectly with the slope changes of the radial stellar velocity, $V$. More interestingly, we observe a reversal of sign in the $h_3$ moment map, creating a ``butterfly'' shape in its central 15\arcsec, away from the thin fast-rotating disc, and up to $\sim$10\arcsec\ above the major-axis. This region of correlated $V$ and $h_3$ may be due to the fact that the LOS intercepts both the spheroidal component, which is dominating the light, and the edges of the fast rotating outer disc, creating a high velocity wing in the LOSVD. Such a correlation is also often generically explained by the orbital structure associated with the combination of a bar~\citep{Athanassoula1999, Chung2004} and an inner-disc, a possibility which may be supported by the double maximum observed in NGC\,3115 rotation curve. The origin of this "butterfly" region in the $h_3$ moment map is further discussed in \S~\ref{subsec:h3}.

Fig.~\ref{Fig:kin} (panel~\textit{d}) shows the $h_4$ map of NGC\,3115, the fourth Gauss-Hermite moment. $h_4$ represents the kurtosis of the LOSVD. A positive (negative) $h_4$ value means that the LOSVD shows a narrower (broader) symmetric profile than a pure Gaussian LOSVD \citep{VanderMarel1993, Gerhard1993, Bender1994}. NGC\,3115 $h_4$ values are rather uniform with positive values of $\sim$0.05, and slightly lower values along the major-axis in the inner 15\arcsec. We also note higher $h_4$ values along the minor-axis, associated with the cone-like structure observed in the dispersion map (with high $\sigma$ values), as well as along the thin stellar disc. Finally, it is important to mention that the shallow gradient we observe in the $h_4$ moment map, from the left to the right side, can likely be associated with the systematic errors noticed in \S~\ref{subsec:kinSystematics}. These systematics are also visible in panel~\textit{(d)} of Figure~\ref{Fig:kin_cut}.

Our results are consistent with the study of \cite{Norris2006}, who already suggested the presence of a fast-rotating, kinematically cold stellar disc component embedded in a slower rotating, kinematically hot spheroid; and \cite{Arnold2011, Arnold2014}, that complements our MUSE study with data extending from 4~--~10\,R$_e$. The MUSE dataset reaches ``only'' about four effective radii, but with obviously much better spatial resolution than any previous study, which is key revealing the full complexity of its kinematic structures.

%%%%%%%%%%%%%%%%%%%%%%%%%%%%%%%%%%%%%%%%%%%%
\section{Stellar population analysis}
\label{sec:pop}

% Figure stellar population
%__________________________
\begin{figure*}
   \centering
   \includegraphics[width=\textwidth]{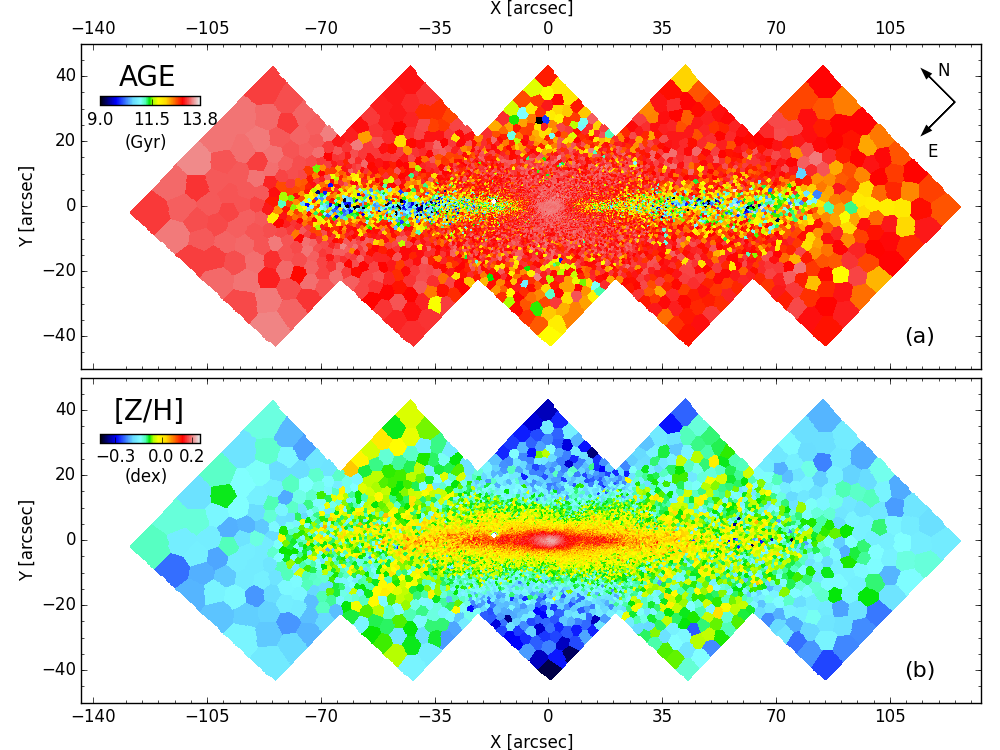}
   \caption{Maps of the stellar population of NGC\,3115 obtained through regularised full spectral fitting: (a) Mass weighted age ; and (b) mass weighted metallicity, [Z/H]. The color code for each panel is indicated on its top left corner. We observe a clear distinction of age between the disc and the spheroid component of NGC\,3115, as well as a clear metallicity gradient along the major- and minor-axis of the stellar disc.}
   \label{Fig:pop}
\end{figure*}

% Figure Age versus V/S
%_____________________
\begin{figure}
   \centering
   \includegraphics[width=\columnwidth]{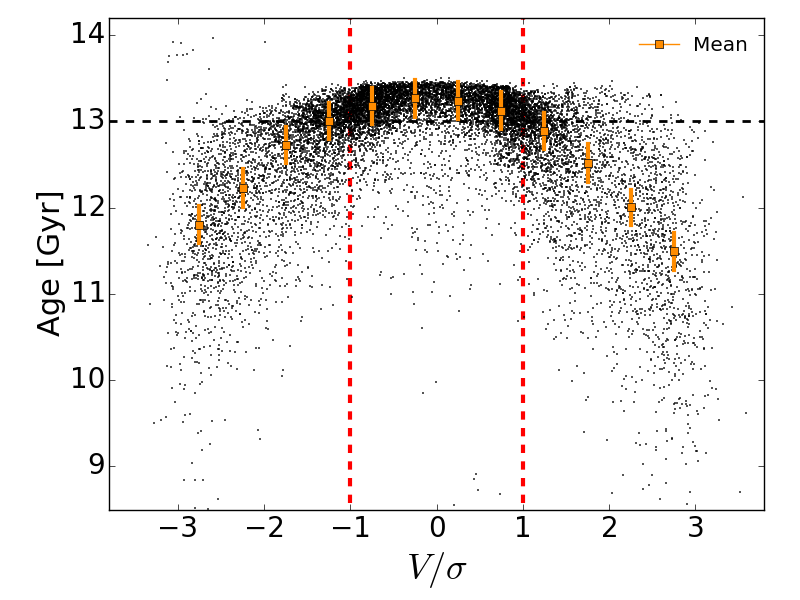}
   \caption{Stellar age (mass-weighted) versus $V/\sigma$, the ratio of the stellar radial velocity and velocity dispersion, for each spaxel of NGC\,3115 MUSE cube. The spaxels corresponding to the fast-rotating disc ($|V/\sigma|$>1) have younger stellar ages (9--13~Gyr) whereas the pressure supported regions of NGC\,3115 ($|V/\sigma|$<1) show a rather uniform and old stellar age distribution ($\geq$13~Gyr). The younger stellar population observed in Figure~\ref{Fig:pop} is therefore well associated to the fast-rotating disc observed in Figure~\ref{Fig:kin}. To help reading the figure, we indicate the age of 13~Gyr by an horizontal dashed line, and plot the mean and standard deviation of the points distribution in ten bins of $V/\sigma$.}
   \label{Fig:Age_VS}
\end{figure}

\subsection{Age and metallicity measurements}
\label{subsec:popMeasure}

To measure the mean stellar age and metallicity of NGC\,3115, we used Single Stellar Populations (SSP) models from the commonly used MIUSCAT library~\citep{Vazdekis2012aMIUSCAT}, which we have previous experience with \citep{Guerou2015}. This library provides single-age, single-metallicity stellar population models theoretically computed from the MILES and CaT \citep{Cenarro2001} empirical spectra~\footnote{We did not use the same libraries for the kinematics and stellar population analyses since the MILES library is optimised for precise radial velocities, therefore preferred for kinematics analysis, whereas the MIUSCAT models reproduce better the continuum and absorption lines shapes, critical for stellar population analysis.}. We used a subset of 120 models covering a age range of 0.5 to 14.1~Gyr (split in thirty bins) and a metallicity range of -0.71 to +0.22~dex (split in four bins), logarithmically sampled and computed using a unimodal IMF with a slope of 1.3, equivalent to a \cite{Salpeter1955} IMF, and based on Padova isochrones \citep{Girardi2000}. These models are included in the ``SAFE'' range of the MIUSCAT models (see \citealt{Vazdekis2012aMIUSCAT} for more details). The models spectra were retrieved from the MILES website\footnote{See http://miles.iac.es/pages/stellar-libraries/miles-library.php} and sampled at a spectral resolution of 2.51\,\AA\,~FWHM.

Although \cite{Norris2006} noticed a significant difference in [$\alpha/Fe$] abundance between the major- and minor-axis of NGC\,3115, we restricted our analysis to use only solar abundance models. While spectral models with variable alpha element abundance ratio are currently available \citep{Conroy2009, Vazdekis2015}, and classic absorption line indices can also be used to infer abundance ratios via models \citep{Thomas1999, Maraston2005, Schiavon2007}, the systematic errors remaining in our data due to non-optimal sky subtraction and observing procedures impede the robust extraction of this higher-order information. We have verified, however, that the star formation timescales implied by our star formation histories are consistent with the abundance ratios measured via absorption line indices and single stellar population models, finding that the disk has lower [$\alpha/Fe$] compared to the spheroidal component (in agreement with \citealt{Norris2006}).

We used this set of models to fit NGC\,3115 spectra using again the pPXF software, following the same methodology as the kinematics analysis (i.e.\,,~same spatial binning, fits performed over the same spectral range, masking of the sky lines, not masking potential galaxy emission lines). However, as recommended by \cite{Cappellari2004ppxf}, we used only multiplicative polynomials (of the 16th order) to account for flux calibration errors, not additive polynomials (which can alter the line strengths and bias the age and metallicity inferred). We apply linear regularisation constraints \citep{Press1992} to the fit via the REGUL keyword in pPXF. As in~\cite{McDermid2015} and \cite{Guerou2015}, we increase the regularisation until the fit to the spectrum becomes unacceptable, defined as when the $\chi^2$ value exceeds that of the unregularised fit by $\sqrt(2N)$, where N is the number of pixels. In this way, we find the smoothest distribution of ages and metallicites that is still statistically consistent with the data. We used the spaxel with the highest S/N to fix the degree of regularization and subsequently kept it constant over the full field.

The quoted age and metallicity values are derived by taking the mean age and metallicity of the templates forming the best fit, weighted by their contribution to it. Each model of the MIUSCAT library is normalised to a star of one solar mass, hence, the quoted age and metallicity values are mass-weighted. Therefore, the weights distribution obtained with pPXF is also the mass distribution of the total stellar mass, projected onto the grid of models, i.e.\,,~the mass per (Age, [Z/H]) combination of each model.

We performed similar analyses with different pPXF settings: masking the potential galaxy emission lines, restricting the fitted wavelength range to 5000--5500\,\AA\,and always recovered the same general structures, both for the age and metallicity maps. We estimate the impact of our fitting parameters to create median absolute differences of $\sim$0.4~Gyr and $\sim$0.05~dex in the determination of the stellar population parameters. Once again, on top of the fitting method, the sky background residuals create systematic errors that are visible in our stellar population maps (i.e.\,,~the shallow gradient from left to right on the age map, Fig.~\ref{Fig:pop}). Assuming symmetric stellar populations, we estimate the age and metallicity systematic errors to be, respectively, 0.5~Gyr and 0.02~dex.

\subsection{Mass-weighted age and metallicity maps from regularised spectral fitting}
\label{subsec:popMaps}

Fig.~\ref{Fig:pop} shows the mass-weighted age (panel \textit{a}) and metallicity maps (panel \textit{b}) obtained from our MUSE reduced cube of NGC\,3115. The first striking result is the clear age distinction between the two main components of NGC\,3115, namely the thin fast-rotating disc and the spheroidal structure. Indeed, we find an age span of 9--13~Gyr for the stars presumably associated with the thin disc, whereas the stars of the spheroid have a rather uniform age of 13--13.5~Gyr. The distribution of the younger stars nicely corresponds to the kinematically cold structure observed in the kinematics (Fig.~\ref{Fig:kin}, panel~\textit{b}), having a radial extent along the major-axis of $\sim$80\arcsec\ and a vertical extent of a few arcseconds at 10\arcsec\ from the centre, to $\sim$15\arcsec\ at 60~--~70\arcsec\ along the major-axis (Fig.~\ref{Fig:pop}, panel \textit{a}). Also noticeable are the central 10\arcsec\,~dominated by very old stars, showing a clear positive gradient toward the centre, going from 12.5~Gyr and peaking at 13.5~Gyr. The association of the younger stellar population (9--13~Gyr) with the prominent fast-rotating disc component is demonstrated in Fig.~\ref{Fig:Age_VS}, where we plot the stellar age distribution of each MUSE spaxel as a function of the local (spaxel-based) $V/\sigma$, the ratio of the stellar mean radial velocity and velocity dispersion. The spaxels corresponding to the fast-rotating disc ($|V/\sigma|$>1) have younger stellar ages (9--13~Gyr) whereas the pressure supported regions of NGC\,3115 ($|V/\sigma|$<1) show a rather uniform and old stellar age distribution ($\geq$13~Gyr). We note that, while the absolute ages of stellar population models are somewhat uncertain, the relative ages between models are more robust \citep{Vazdekis1999}, meaning that the 2-3 Gyr decrease in the mass-weighted mean age of the disc is secure.

Regarding the stellar metallicity, Fig.~\ref{Fig:pop}~(panel~\textit{b}) shows that the disc component exhibits a significant gradient along its major-axis, having a metal-rich core of $\sim$~+0.2~dex, and a low metallicity at its extremity ($\sim$~-0.1~dex at 80\arcsec). The gradient is quite strong in the central region, dropping from +0.2~dex at the centre to +0.1~dex at 10\arcsec\,,~and shallower at larger radii. NGC\,3115 shows also a strong negative metallicity gradient along its minor-axis, from +0.2~dex at its centre to $\sim$~-0.2~dex at 20\arcsec\,. At very large radii, the spheroid component shows a uniform metallicity distribution of $\sim$~-0.15~dex. Other interesting features are visible in the metallicity map of NGC\,3115, such as a significantly lower metallicity of the spheroid component along the minor-axis (down to $\sim$~-0.25~dex) and flaring-like structures (with higher metallicity values of $\sim$~-0.05~dex) at $\sim$40\arcsec\,~along the major-axis. Unfortunately, it is hard to interpret these structures since they seem to correspond (spatially) to the different MUSE exposures. They could be due to real differences in NGC\,3115 stellar populations or alternatively could be artifacts of our treatment of the sky variation. However, we can confirm that the stellar (flattened) spheroid is made of significantly more metal-poor stars than the disc component, the latter also showing strong gradients along both its radial and vertical extent.

While these results are in agreement with and confirm what was inferred by~\cite{Norris2006}, the MUSE data allow us to unambiguously distinguish the different stellar population parameters of the spheroidal and disc of NGC\,3115.

% Figure stellar population: Mass Fraction
%________________________________________
\begin{figure*}
   \centering
   \includegraphics[width=\textwidth]{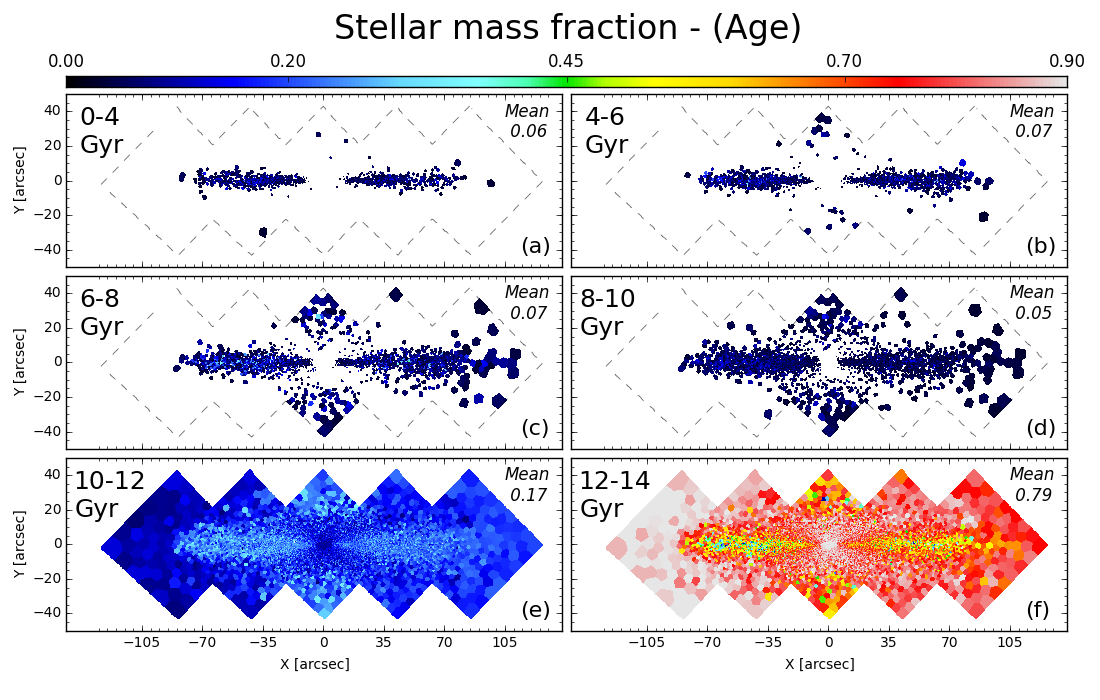}
   \caption{Stellar mass fraction maps of NGC\,3115, in six bins of age, obtained by projecting the stellar models weighting distribution solution, obtained with pPXF, onto the grid parameters (Age, [Z/H]). For each panel, the age bin limits are indicated on its top-left corner, the mean stellar mass fraction on its top-right corner, and the color scheme by the color bar at the top of the figure. Spaxels containing a stellar mass fraction lower than 0.03 are masked, considered below our uncertainties. The galaxy orientation is similar to Fig.~\ref{Fig:kin}. The extended star formation history of the outer disc component is clearly visible in the first four panels (\textit{a} to \textit{d}), whereas the stars present in the spheroid were formed early (age$\geq$10~Gyr), with more than 85\% of its stellar mass formed between 12~--~14~Gyr (panels \textit{e} and \textit{f}).}
   \label{Fig:popMassVSage}
\end{figure*}
 
% Figure stellar population: Mass maps 
%__________________________________________
\begin{figure*}
   \centering
   \includegraphics[width=\textwidth]{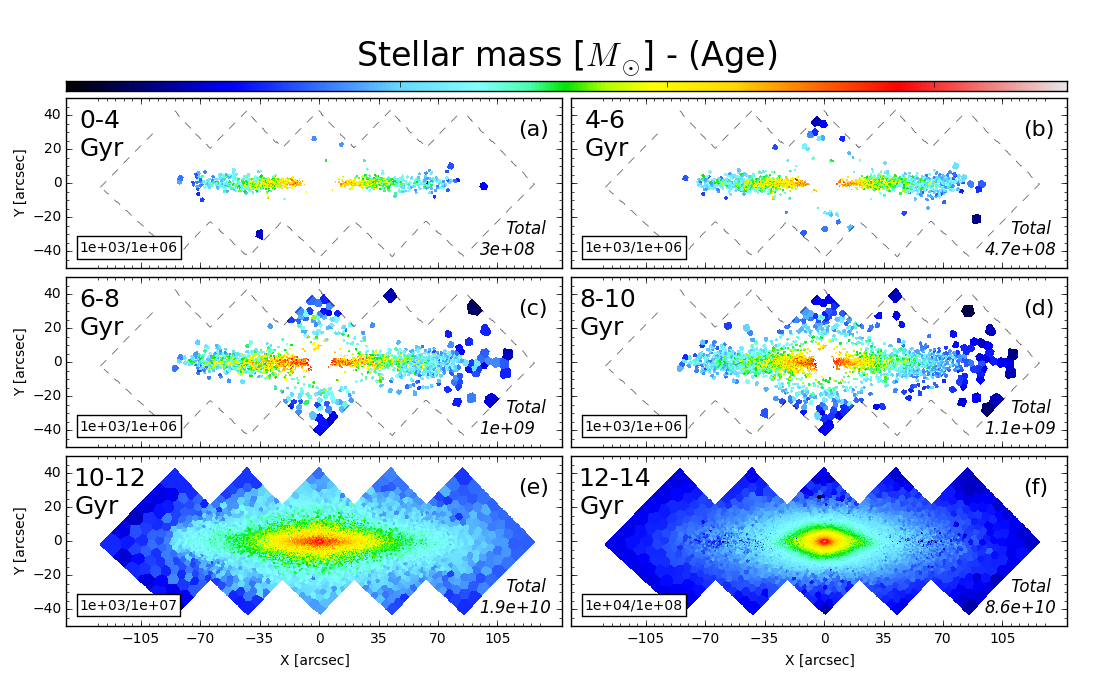}
   \caption{Stellar mass maps of NGC\,3115, in six bins of age, as in Figure~\ref{Fig:popMassVSage}. These maps show the present day stellar mass distribution as a function of its mass-weighted age. For each panel, the total stellar mass per bin of age (in solar mass) is indicated at its bottom-right corner, and the color scheme limits (corresponding to the color bar at the top) are indicated on its bottom-left corner, and vary from one another. The noticeable change in the shape of the iso-mass contours between 10~--~12~Gyr and 12~--~14~Gyr, from rather flat to more spheroidal, suggests a more dissipative mass assembly at early epoch (age>12~Gyr) possibly from a major merger. Panels \textit{(a)}~to~\textit{(c)} suggest that extended star formation occurred in-situ in the disc component of NGC\,3115.} 
   \label{Fig:MassMaps}
\end{figure*}

% Figure stellar population: mass Fraction VS Z
%_______________________________________________    
\begin{figure*}
   \centering
   \includegraphics[width=\textwidth]{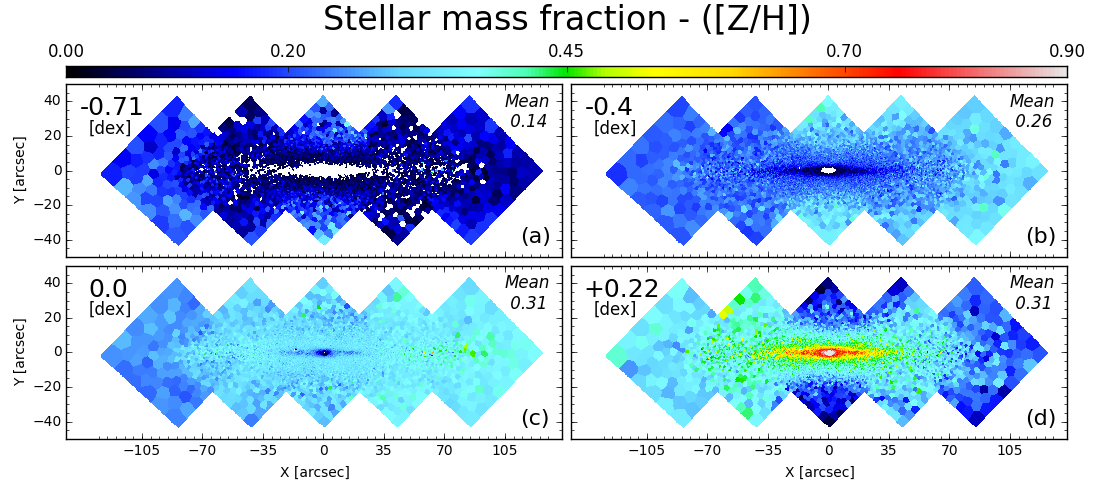}
   \caption{Stellar mass fraction maps of NGC\,3115, in four bins of metallicity, marginalised over age. For each panel, the metallicity bin limits are indicated on its top-left corner, the mean stellar mass fraction on its top-right corner, and the color scheme by the color bar at the top of the figure. Spaxels containing a stellar mass fraction lower than 0.03 are masked, considered below our uncertainties. The galaxy orientation is the same as Fig.~\ref{Fig:kin}. It is remarkable that the central 10\arcsec\ where the inner disc resides, are made of only metal-rich stars (panel \textit{d}). The disc component has very little metal-poor stars (panels \textit{a} and \textit{b}), and the spheroid of NGC\,3115 has a flat distribution over the four metallicity bins, from metal-rich to metal-poor.}  
   \label{Fig:popMassVZ}
\end{figure*}

% Figure Mass Fraction maps of old/young VS rich/poor
%_____________________________________________________
\begin{figure*}
   \centering
   \includegraphics[width=\textwidth]{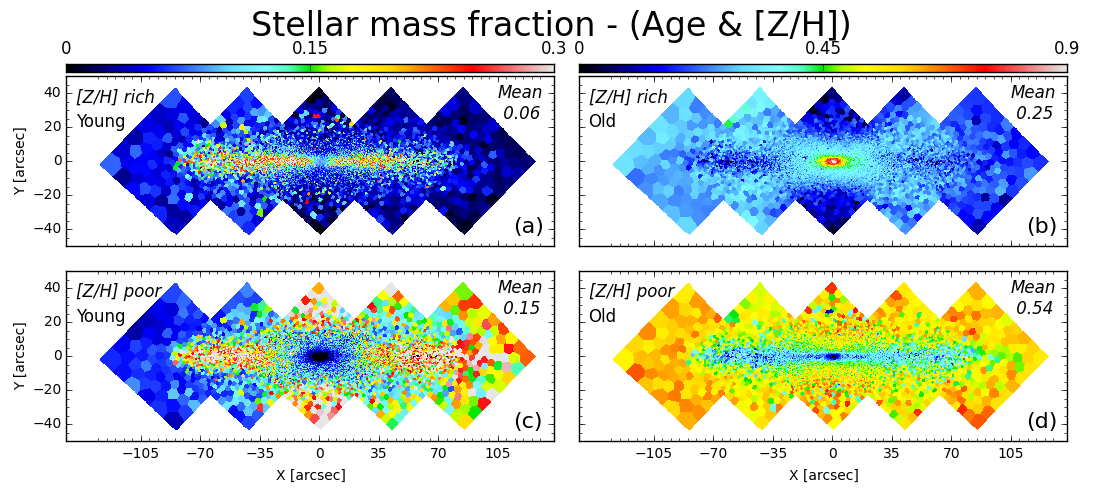}
   \caption{Stellar mass fraction of NGC\,3115 in four bins of age/metallicity pairs: young (age$\lesssim$12~Gyr), old (age>12~Gyr), metal-rich ([Z/H]= 0.2~dex) and metal-poor ([Z/H]<0.2~dex). For each panel, the age and metallicity bin limits are indicated on its top-left corner, the mean stellar mass fraction on its top-right corner, and the color scheme by the color bar at the top of its respective column. The old and metal-poor stellar population of NGC\,3115 is mostly located in its spheroid component, whereas the young and metal-rich stars are mostly in the disc component.}
   \label{Fig:ComponentPopFrac}
\end{figure*}

\subsection{Star formation history and mass assembly of NGC\,3115}
\label{subsec:sfh}

We used our stellar population analysis to further look at the star formation history of NGC\,3115. As explained in \S\,\ref{subsec:popMeasure}, the weights distribution obtained with pPXF is the mass distribution of the total stellar mass of the galaxy, i.e.\,,~the mass per (Age, [Z/H]) combination of each model. We can thus reconstruct a two-dimensional spatial view of the star formation history of the galaxy, by taking the weighting distribution of the stellar models forming the best fit, in each spaxel of the IFU observations. 

Here we present two ways of looking at this reconstructed star formation history: first, by showing the mass fraction (as compared to the present day total stellar mass), of each stellar model (grouped in bins of age and/or metallicity); and second, by reconstructing the total stellar mass of the galaxy in different bins of age. We warn the reader that these maps do not reconstruct at face value the mass assembly history of the galaxy, but rather show the present day stellar mass distribution (model dependent) as a function of its mass-weighted age and/or metallicity, assuming a smooth star formation history. Nonetheless, these maps can help constraining scenarios for the formation and evolution of NGC\,3115.

Fig.~\ref{Fig:popMassVSage} and Fig.~\ref{Fig:MassMaps} show the stellar mass fraction and stellar mass, respectively, in six bins of age: from 0~--~4~Gyr, 4~--~6~Gyr, and up to 14~Gyr in bins of 2~Gyrs. From Fig.~\ref{Fig:MassMaps}, we derive a total stellar mass of $1.1\times10^{11}$~\Msun\ for NGC\,3115, within the field of view covered by MUSE, consistent with previous estimates \citep{Karachentsev2014}. Most of this stellar mass ($\sim$95\%, i.e.\,, $1.05\times10^{11}$~\Msun,) is made of old stars (>10~Gyr), according to panels \textit{(e)-(f)} of Fig.~\ref{Fig:popMassVSage} and Fig.~\ref{Fig:MassMaps}. This old stellar population represents the entire stellar content of the spheroid component, with $\sim$85\%\,~of its associated stellar mass formed very early, at a redshift \textit{z}$\sim$3 and greater (12~--~14~Gyr ago), and the remaining $\sim$15\% formed only soon after (10~--~12~Gyr ago). The central region ($\sim$10\arcsec) is also made of old stars only (older than 10~Gyr) which suggests a lack of significant (by mass) gas infall, followed by star formation during the past 10~Gyr at the very centre of NGC\,3115. From a more general perspective, Fig.~\ref{Fig:MassMaps} (panel \textit{e}) shows a more flattened mass distribution for the stellar population with ages between 10 and 12~Gyr, than the oldest stellar population (12~--~14~Gyr). This suggests that dissipation may have played an important role in the formation and assembly of the central region of NGC\,3115 at early epoch (>12~Gyr, i.e.\,, $\textit{z}\gtrsim3$). In contrast, the outer stellar disc component has only $\sim$50\% of its total stellar mass formed between 12~--~14~Gy, and $\sim$20\% between 10~--~12~Gyr.  

The first four panels of Fig.~\ref{Fig:popMassVSage} and Fig.~\ref{Fig:MassMaps}, namely \textit{(a)-(b)-(c)-(d)}, clearly show that star formation persisted in the outer stellar disc between \textit{z}$\sim$2 and the present day, whereas the spheroid component and the central regions ($\sim$10\arcsec) do not contain stars younger than 10~Gyr. Between $\sim$20~--~30\% ($5.7-8.4\times10^{8}$~\Msun) of the total outer disc stellar mass was formed between 0~--~10~Gyr, with a rather uniform distribution in each bin of age ($\sim$5~--~7\%), as shown in Fig.~\ref{Fig:popMassVSage}. From Fig.~\ref{Fig:MassMaps}, we derive a mean star formation rate (SFR) in the disc of NGC\,3115 of 0.4~\Mperyr\ between 10~--~6~Gyr, decreasing to 0.2 and 0.08~\Mperyr, between 6~--~4~Gyr and 4~--~0~Gyr, respectively. This nicely suggests that no major-mergers occurred over the last 10~Gyr (as mentioned by \citealt{Norris2006}) as this would certainly have destroyed such a well defined stellar disc and disrupted such a steadily decreasing star formation rate.

Figure~\ref{Fig:popMassVZ} shows the stellar mass fraction in four metallicity bins:~$-0.7$~dex ; $-0.4$~dex ; $0.0$~dex ; and $+0.22$~dex, where we see again the clear distinction between the stellar disc of NGC\,3115 and its spheroid component. Indeed, $\sim$40~--~80\% of the disc is composed of metal-rich stars (panel \textit{d}), $\sim$20\% have solar-metallicity (panel~\textit{c}), and less than $\sim$10\% of its stellar mass is found to be metal-poor (panels \textit{a-b}), with no contribution at all of the most metal poor bin ($-0.7$~dex) to the inner 40\arcsec of the disc. In contrast, the spheroid component has a rather uniform and similar distribution in the four metallicity bins (within our uncertainties), each one containing $\sim$20~--~30\% of its total stellar mass. This is very interesting as it tells us that the spheroid component has (on average) solar metallicity and lower but still contains a significant fraction of metal-rich stars. It is also interesting to notice that the central part of NGC\,3115 (including the inner disc) and up to $\sim$40\arcsec\,~along the major-axis, is only made of metal-rich stars (i.e., we do not detect a significant metal-poor component), naturally leading to the observed increasing gradient in the central 10\arcsec (see \S\,~\ref{subsec:popMaps}). This steep gradient suggests that the star formation of the disc components happened in-situ~\citep{Kobayashi2004, Pipino2010}, in agreement with their extended star formation histories. 

Finally, we also analyse the stellar mass fraction distribution of NGC\,3115 as a function of the combined stellar age and metallicity (Fig.~\ref{Fig:ComponentPopFrac}), for four pairs of age/metallicity bins. These four bins have been selected based on Fig.~\ref{Fig:popMassVSage}~and Fig.~\ref{Fig:MassMaps}, in order to optimise the separation of the disc and spheroid components. We chose to split the stellar populations with ``young'' and ``old'' ages ($\lesssim$12~Gyr, and $>$12~Gyr, respectively), and with ``metal-rich'' and ``metal-poor'' components ([Z/H]=0.2~dex, and [Z/H]$\lesssim$0.2~dex, respectively). Panels \textit{(b)-(d)} illustrate well that the central regions and the spheroid of NGC\,3115 are mostly made of old stars, metal-rich in the core of the galaxy, metal-poor ([Z/H]<+0.2~dex) in the spheroid. The outer disc is made of (on average) $\sim$50\% of old stellar population, uniformly distributed between the metal-rich and metal-poor bins. Regarding the young stellar population (panels \textit{a-c}), most of its mass is distributed in the stellar disc, but in a different manner between the metal-rich and metal-poor stars. Indeed the metal-rich ([Z/H]=0.2~dex) young stellar population is present all along the stellar disc, from its extremity (at $\sim$80\arcsec), to the core of NGC\,3115 ($\sim$5\arcsec\,from the centre) and having a rather thin and uniform vertical extent ($\sim$5~--~10\arcsec). On the other hand, the metal-poor ([Z/H]<0.2~dex), young stellar population has a thicker vertical extent ($\sim$20\arcsec) and located on the external part of the outer disc, i.e.\,, between 20\arcsec\,and 80\arcsec\,along the major-axis of NGC\,3115. It would be tempting to associate this distinction in stellar population with the observed kinematics which suggest a thicker disc component (see \S\,~\ref{subsec:kin}), but this is beyond what we can confirm with the present analysis. 

Through these star formation history and mass assembly analyses, we provide a striking illustration of the potential of MUSE spectroscopic data to reveal exquisite details of the stellar population distribution. The formation and evolution scenarios of the different components of NGC\,3115 are further discussed in the following section \S\,~\ref{sec:discussion}.

%%%%%%%%%%%%%%%%%%%%%%%%%%%%%%%%%%%%%%%%%%%%
\section{Discussion}
\label{sec:discussion}

\subsection{NGC\,3115, an S0 galaxy with discs, spirals and rings.}
\label{subsec:substructures}

\begin{figure}
   \centering
   \includegraphics[width=\columnwidth]{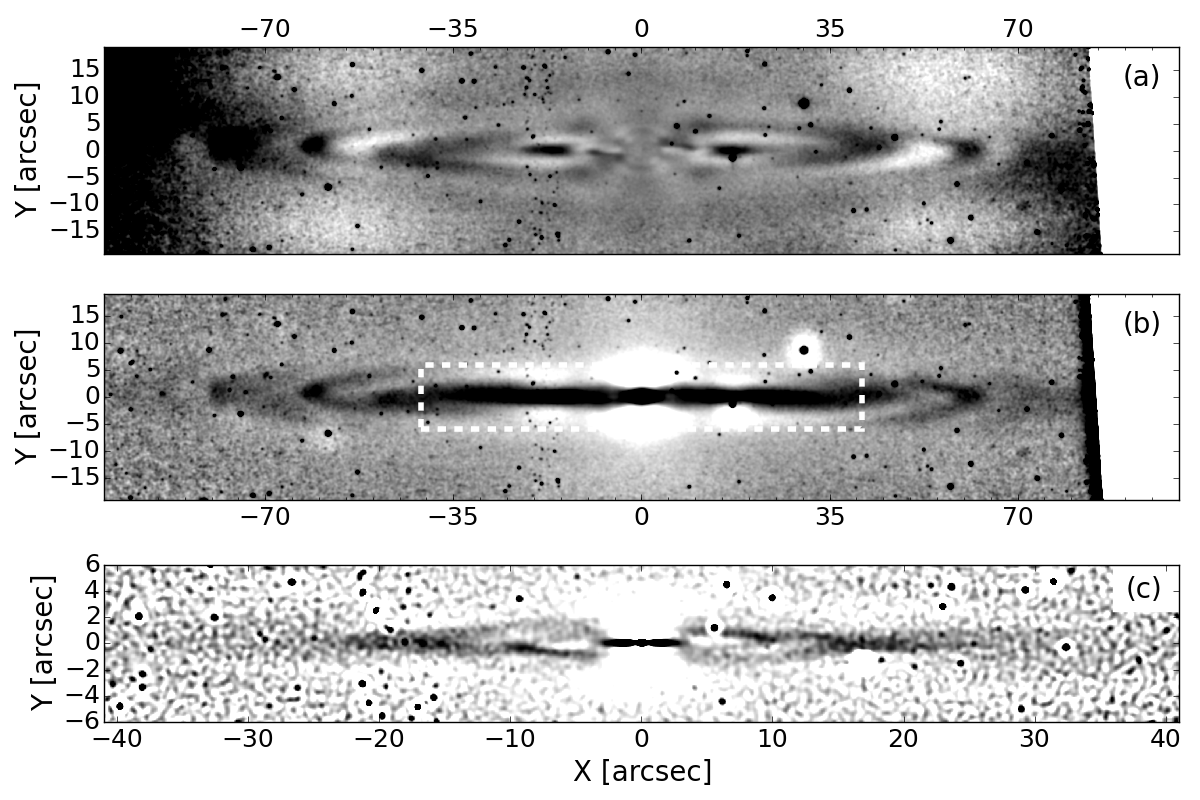}
   \caption{\textit{(a)}: ACS/HST residuals after removal of the MGE photometric model~\protect\citep{Emsellem1999}; \textit{(b)}: Unsharp masking of the HST image; \textit{(c)}: Zoom corresponding to the dashed rectangle of the unsharp masking image (panel \textit{b}).}
   \label{Fig:residuals}
\end{figure}

Based on photometry alone, NGC\,3115 is classified as an S0 galaxy. Its flattened spheroid component dominates the light, even though the outer and inner disc are very prominent due to the near edge-on inclination. Going beyond this simple description, NGC\,3115 has a lot of substructures which deserve to be discussed. Fig.~\ref{Fig:residuals} shows the high resolution photometric residuals (panel~\textit{a}) of NGC\,3115 obtained by subtracting the photometric model \citep{Emsellem1999} from an ACS/HST F475W image (PI IRWIN, ID-12759). Clear signatures of spiral- or ring-like structures are detected in this residual image, confirming what we observed using the reconstructed MUSE V-band image~(see Fig.~\ref{Fig:phot_res}, panel~\textit{c}). In the outer part ($R > R_e$), three main features are identified on each side of the galaxy, with characteristic radii of $\sim$40\arcsec, $\sim$60\arcsec\,and $\sim$75\arcsec\,respectively from the galaxy centre. We can rule out that these structures are model dependent artifacts from the model subtraction as they appear both in the unsharp-masked and model-subtracted images (see panel~\textit{b} of Fig.~\ref{Fig:residuals}).

% residuals from HST images
%__________________________

In the central region, this unsharp masking technique also beautifully emphasises the presence of the very thin (less than $\sim$0$\farcs5$ in height) nuclear-disc of 3\arcsec\ in radius (see Fig.~\ref{Fig:residuals}, panel~\textit{c}), first detected by \cite{Lauer1995} and \cite{Kormendy96}. This nuclear disc spatially matches the steep gradient observed in the central 3\arcsec\ of NGC\,3115 rotation curve as well as the inner $h_3$ rise (see Fig.~\ref{Fig:kin_cut}, panels \textit{a} and \textit{c}). Additionally, we observe two anti-symmetric substructures at a radius of $\sim$10\arcsec\ and extending down to 4\arcsec. As already suggested by \cite{Emsellem2002a}, these structures (partially hidden by the bright nucleus) might extend, respectively, above and below the nuclear disk to connect on its opposite edges, thus interpreted as two inner spiral-arms (see Fig.~2 of \citealt{Emsellem2002a}). Further away, at $\sim$15--20\arcsec\,the residuals suggest a pair of (inner-) spiral arms or rings, also possibly associated with the corresponding plateau in the stellar velocity profile and the abrupt changes in the velocity dispersion, $h_3$ and $h_4$ profiles. Although these structures are again present in the HST residual image, it is hard to strictly define their nature and exact (deprojected) morphology. The presence of a nuclear ring in NGC\,3115 was already reported by \cite{Savorgnan2015}, quoting a radial extent of 15\arcsec\ for this component.

\begin{figure*}
   \centering
   \includegraphics[width=\textwidth]{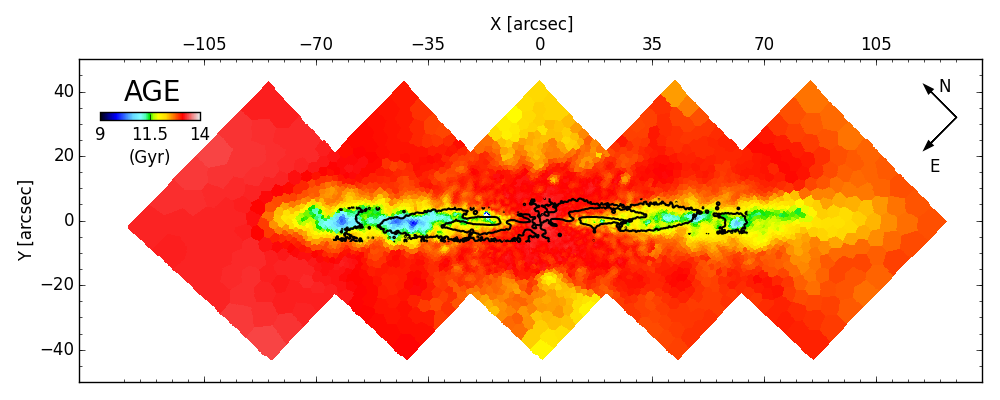}
   \caption{Smoothed version of the stellar age map of NGC\,3115, with the ACS/HST residuals contours over-plotted on top. We plotted only the zero-level contours corresponding to the potential spiral-arms in order to enlighten their boundaries. A clear spatial match is observed between the spiral-like residual structures and the youngest stellar population observed.}
   \label{Fig:spirals}
\end{figure*}

The existence of the outer spiral-like features have been already suggested by several authors in the literature~\citep{Capaccioli1987, Norris2006, Michard2007}, always from a photometric point of view. Here, we report for the first time their detection based on the stellar populations derived from spectroscopy. In Fig.~\ref{Fig:spirals}, we mark the spiral-like structure using contours from the HST residual image on top of the (smoothed) mean stellar age map of NGC\,3115. Remarkably, the photometric spiral-like features roughly match the younger part of what we so far interpreted as a smooth stellar disc. The inter-arm regions show, on average, stars that are older than the presumed spiral-arms, something which is observed on both sides of the galaxy. We emphasise here that the age difference noticed between the spiral-like structures and the inter-arms may simply be due to the contrast between these two regions: the spiral arms are mass over-densities, relatively to the inter-arms regions. If the stars in the disc are younger overall, this increases the contrast with the presumed older background and foreground stars, hence the younger mass-weighted age. In other words, what we detect here are primarily mass over-densities, and not necessarily local age variations in the disc itself. In the inner region, we do not detect any specific features associated with the presumed inner spiral arms. This is no proof of an homogeneous stellar population, as actual stellar population differences of such low contrast and high spatial frequency features are inevitably strongly diluted relative to the dominant metal-rich, older central stellar components of NGC\,3115. Looking at possible signatures of the outer overdensities in the individual age bins on Fig.~\ref{Fig:popMassVSage}, we do detect significant increases in the mass fraction of young stars (for ages from 0 to 10~Gr) specifically between 15\arcsec\ and 20\arcsec. There are also small peaks near 40\arcsec\ and 60\arcsec\ although these are only marginally detected. Interestingly, the increase near a radius of 40\arcsec\ is shifted away from the major-axis following the anti-symmetric features observed in Fig.~\ref{Fig:residuals}.

One open question that still persists is about the formation time and development of such spiral structures in NGC\,3115. The age distribution observed in Fig.~\ref{Fig:spirals} only indicates the mean (mass-weighted) stellar age of the population forming these structures. At a radius of 50\arcsec\ ($\sim 2.5$~kpc), the timescale for a circular orbit is about 60~Myr. It is thus unlikely that the observed spiral-like structures have been frozen (evolution free) over the last $\sim$9~--~10~Gyr, as spiral structures have typical lifetimes of two to three orbital revolutions~\citep{Sellwood2010}. Still, such density waves may persist (and evolve) in a non self-gravitating disc, as long as the wave can propagate and the system does not get too (dynamically) hot, favored by the presence of gas (see \citealt{Bekki2002}). However, we cannot distinguish that scenario from the one where the spirals recently develop out of the - on-average younger - disc material. Such density waves could also be imposed via coupling to a tumbling bar, the presence of which we briefly discuss in the following section \S\,~\ref{subsec:h3}.

% Spirals evidence from populations
%__________________________________

%_______________________________________
\subsection{Signs of a bar in NGC\,3115}
\label{subsec:h3}

In addition to the complex structures discussed in \S\,~\ref{subsec:substructures}, there are, in the stellar kinematics, signatures that could be associated with the presence of a bar in the central region of NGC\,3115. \cite{Athanassoula1999} performed simulations of peanut-bulge dominated galaxies and concluded that in galaxies seen edge-on, strong bars may show up as:~\textit{(i)} a double maximum in the stellar rotation profile along the major-axis, \textit{(ii)} a flat central stellar velocity dispersion profile (sometimes with a dip), \textit{(iii)} an $h_3$ Gauss-Hermite moment correlated with the stellar radial velocity, $V$, over the bar length, and \textit{(iv)} a $h_3$ Gauss-Hermite moment anti-correlated with $V$ in the very centre where nuclear-discs are often present. Such signatures have been observed by \cite{Chung2004} in edge-on, bulge dominated spiral galaxies, and more recently, \cite{Iannuzzi2015} discussed the dependence on inclination, strength of the bar/peanut components, and outlined features which may be observed with integral-field spectroscopy.

\begin{figure}
   \centering
   \includegraphics[width=\columnwidth]{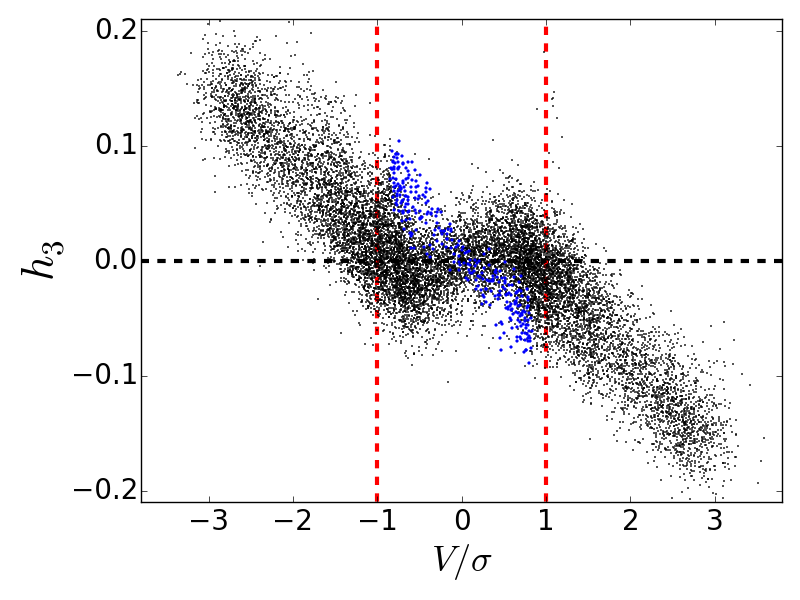}
   \caption{Third Gauss-hermite moment, $h_3$, versus $V/\sigma$ for each spaxel of NGC\,3115 MUSE cube. Blue dots represent the spaxels corresponding to the nuclear-disc (central 3\arcsec\,$\times$\,1\arcsec) of NGC\,3115.}
   \label{Fig:longChair}
\end{figure}

Except for the flat central velocity dispersion (point \textit{ii}), we observe all signatures mentioned above in NGC\,3115. The most striking signature is shown in Fig.~\ref{Fig:longChair}, where we plot the third Gauss-Hermite moment of the LOSVD, $h_3$, as a function of the local ratio $V/\sigma$ for each spaxel of the covered MUSE field of view. A strong anti-correlation is visible for ${|V/\sigma|\gtrsim}$1, whereas $h_3$ is mostly correlated with $V/\sigma$ for ${|V/\sigma|\lesssim}$1. More interestingly, we also plot the spaxels corresponding (spatially) to the nuclear-disc of NGC\,3115~(Fig.~\ref{Fig:longChair}, blue dots) and observe a strong anti-correlation, in agreement with the simulations predictions (point \textit{iv}). The bar orientation, if seen almost end-on, or the presence of a central dark mass may be partly responsible for the fact that we do not observe a flat velocity dispersion profile in NGC\,3115.

% Figure h3 versus V/S
%_____________________
 
% Figure models MAPS
%_____________________
\begin{figure}
   \centering
   \includegraphics[width=\columnwidth]{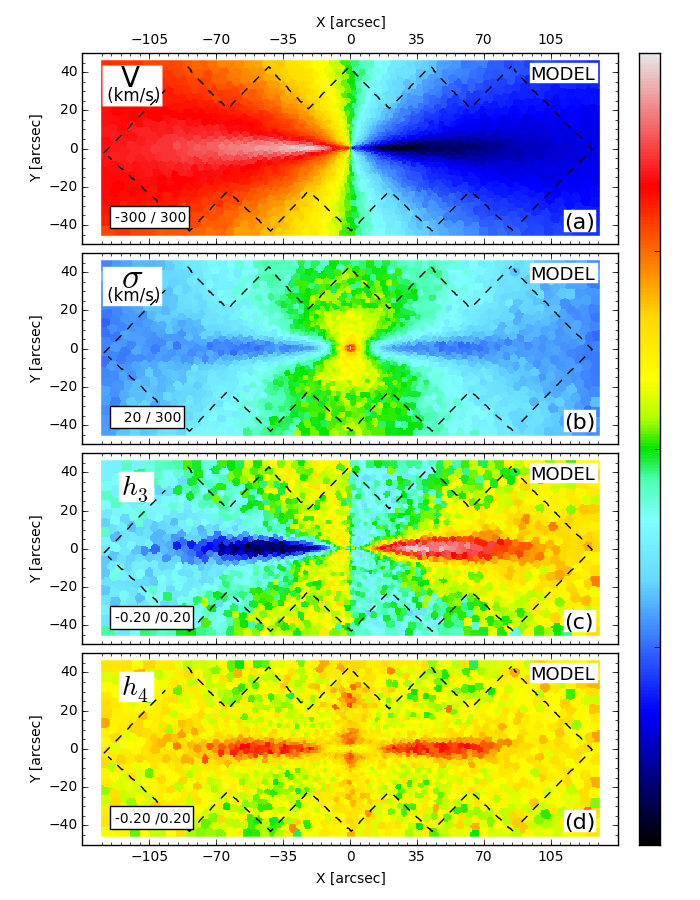}
   \caption{As in Fig.~\ref{Fig:kin}, but for the dynamical model described in \S\,~\ref{subsec:h3}. The dashed contours indicate the MUSE poitings coverage.}
   \label{Fig:modelsMaps}
\end{figure}

To further constrain whether or not a bar could be responsible for such kinematics signatures, we built a simple dynamical axisymmetric model and compared the obtained LOSVD to the one observed with MUSE. This model is based on the Multi-Gaussian Expansion formalism \citep{Emsellem1994} and follows the procedure as sketched in \cite{Emsellem2013}. The MGE model is used to obtain the three-dimensional light distribution, that combined with mass-to-light ratios (M/L) and the use of a dark matter halo as in \cite{Cappellari2015}, leads to an analytic description of the three-dimensional mass distribution of the galaxy. The M/L of each Gaussian were chosen as roughly constant and consistent with the published model of \cite{Emsellem1999}. We implemented some typical variations of the order of 20\%, with a slight decrease of the M/L for the younger (disc) components as revealed by the age map, and a higher M/L for the most outer regions, as to follow the outer $V^2+\sigma^2$ map. Note that the choice of these individual M/L is highly degenerate and was not meant as to exactly fit the details of the observed kinematics. A full-fledged fitting algorithm would be needed to truly optimise the combined anisotropy\,--\,M/L\,--\,dark matter parameters, but this is clearly beyond the reach of the present paper. We make a realisation of that model using particles, fully consistent with the given mass distribution, and solve the Jeans Equations in cylindrical coordinates assuming individual anisotropy parameters for each Gaussian, thus fixing the components of the dispersion tensor ($\sigma_R$, $\sigma_{\theta}$, $\sigma_z$). The resulting N-body model is further projected using an inclination angle of $84.5\degr$, and LOSVDs are reconstructed on a spatial two-dimensional grid within the field covered by the MUSE pointings. These LOSVDs are finally fitted individually with Gauss-Hermite functions, providing individual measurement, for each point of the grid, of the mean radial velocity $V$, velocity dispersion $\sigma$, and higher order Gauss-Hermite moments, $h_3$ and $h_4$ \citep[as in e.g.\,,][]{Emsellem2013}.

We have tuned the anisotropy parameters so that the projected moment maps qualitatively ressemble NGC\,3115's observed stellar kinematics. However, it is important to note that this model is not meant as a fit, in contrast with what is done when applying e.g., a Schwarzschild modelling procedure. Still, we manage to reproduce (see Fig.~\ref{Fig:modelsMaps}) most of the observed structures, namely: the double maximum in the velocity profile along the major-axis, the overall dispersion map with the central high dispersion region extending along the minor-axis, the $h_3$ versus $V$ anti-correlation along the major-axis, as well as the amplitude of each moment over most of the field. The most surprising result comes from the correlation between $h_3$ and $V$ which naturally emerges from the MGE model in the central 10\arcsec\ away from the major-axis, which is very similar both in its extent and amplitude to the structure seen in the $h_3$ MUSE map. Remarkably, we expected this structure to provide a strong hint for the presence of a tumbling bar, as mentioned above, but this dynamical MGE model tells us that an axisymmetric model, following the light distribution of NGC\,3115, can also reproduce this feature (with the help of the fast rotating outer stellar component). This means that we cannot use the observed $h_3$-$V$ correlation as a strong indication that a bar is present in NGC\,3115. We nevertheless witness a complex set of embedded structures with characteristic radii: inner and outer discs with radial extents of 3\arcsec\ and $\sim$85\arcsec\ respectively, spiral-like features at 10\arcsec\,, 40\arcsec\ and 60\arcsec\ in radius, all reflected in local or gradual changes in the observed photometry (Fig.~\ref{Fig:residuals}), kinematics (Figs.~\ref{Fig:kin} and \ref{Fig:kin_cut}), and (partly) stellar populations (Figs.~\ref{Fig:popMassVSage} and \ref{Fig:spirals}). This is strongly reminiscent of a bar-driven evolution, which would indeed tend to lead to differentiation of properties such as mass, disciness and/or dynamical support, or metal distribution. 

Barred galaxies are associated with specific orbital structures, together with regions localised by dynamical resonances \citep{Sanders1976, Schwarz1984, Sellwood1993}. An inner disc is thus often the result of reshaping of the central regions inside the inner Lindblad resonance(s) (ILR, 2:1 resonance), gas fueling and subsequent star formation. Such a disc is visible as a first bump in the stellar kinematic profile (e.g., velocity and velocity dispersion), but can also be emphasised by further metal enrichment or connecting spiral arms \citep{Cole2014}. The bar itself often ends close to the so-called ultra-harmonic resonance (UHR, 4:1 resonance), well inside the co-rotation radius (CR), again associated with spiral arms and sometimes a resonant (squarish) ring \citep{Buta1996,Buta1999}. An outer disc may extend beyond the CR reaching out close to an outer oval ring at the outer Lindblad resonance (OLR, 2:1 resonance). 

\begin{figure}
   \centering
   \includegraphics[width=\columnwidth]{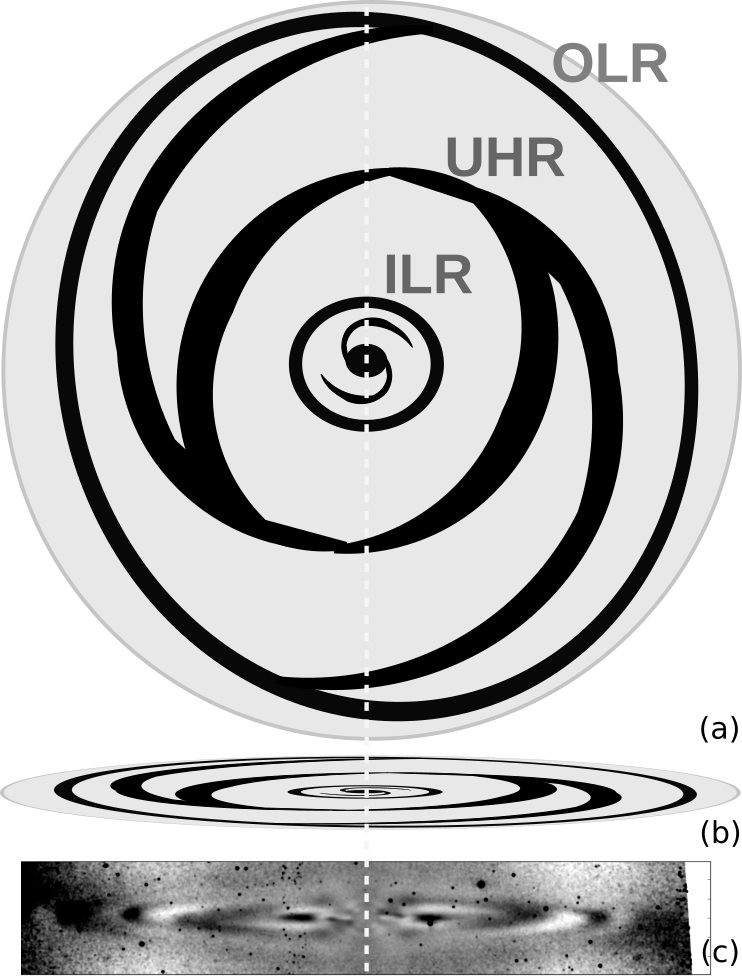}
   \caption{\textit{(a)} Deprojected and, \textit{(b)} projected sketches of the presumed spiral and ring-like features observed in NGC\,3115. \textit{(c)} HST residual image as in Fig.\ref{Fig:residuals}, panel \textit{(a)}.}
   \label{Fig:sketch_n3115}
\end{figure}

How relevant are such generic features to the ones we observe in NGC\,3115? Using the mass MGE model, we estimated the (axisymmetrised) location of such bar-driven resonances by just computing the linear approximation for $\Omega$, the circular frequency, and $\kappa$ the first radial epicycle frequency. By assuming that the inner ring at R=$15-20$\arcsec\ is the ILR,  we can estimate the locations of the UHR, CR, and OLR. To account for uncertaintities in the mass model, as well as for the fact that this is an axisymmetric approximation, we also used a different MGE model where we fixed the local mass-to-light ratios of all Gaussians. We then get radii for the UHR, CR, and OLR of $35 \pm5$\arcsec, $50\pm7$\arcsec\ and $80\pm10$\arcsec. The feature at $\sim 40$\arcsec\ (Sect.~\ref{subsec:substructures}) could thus be associated with an UHR ring, often marking the end of the bar. The extent of such a bar would nicely correlate with the extent of metal-rich stars (Fig.~\ref{Fig:popMassVZ}, panel \textit{c}). The location of the OLR would also roughly coincide with the outer overdensity at 75\arcsec, also close to the end of the disc. A sketch of this scenario is provided in Fig.~\ref{Fig:sketch_n3115}, where we present the deprojected versus the projected residual maps as in Fig.~\ref{Fig:residuals} (a sketch of the presumed inner spiral was presented in \citealt{Emsellem2002a}).

% Figure SKETCH
%_____________________

Whether this picture is correct or not, the observed embedded structures suggest that NGC\,3115 has been significantly re-shaped by a bar, although it does not tell us whether the bar itself is still present. To prove the presence of that bar would require a much more elaborate photometric and kinematic modeling, or new observational signatures. Although having a bar in such a discy early-type galaxy is not necessarily a surprise, it is an important ingredient to keep in mind when trying to constrain the last 10~Gyr of NGC\,3115 history, as this may have been a significant driver for the spatial distribution of metals.

\subsection{Formation and evolution scenarios for NGC\,3115}
\label{subsec:formation_3115}

To explain the formation and evolution of early-type galaxies (ETGs), the so-called ``two phase'' scenario of \citealt{Oser2010} has been quite extensively used and discussed \citep[][and references therein]{Arnold2014, Pastorello2015}, and suggested for NGC\,3115 by \cite{Arnold2011}. According to this scenario, ETGs growth would first be dominated by in-situ star formation at high-redshift ($z\gtrsim2-3$), triggered by dissipative gas collapse forming most of the central stellar mass. In a second phase, at intermediate to low redshifts ($z\lesssim2-3$), ETGs would mostly grow from accretion of stars, via minor mergers, that would populate the external part of galaxies. This difference of mass growth would be marked by a transition radius where the stellar kinematics and population properties would change noticeably. 

From the mass assembly described in \S\,~\ref{subsec:sfh}, we can infer that the stellar mass growth of most of the central region and disc components have been dominated by in-situ star formation: at early epochs ($t\gtrsim12$~Gyr, i.e.\,, $z\gtrsim2-3$) for its central region (central 15\arcsec), and until a recent past (within the last 4~Gyr) for the stellar disc.
The origin of the fast rotating and flattened spheroid ($\sim$200\,\kms\, at 3 $R_e$) is still unclear, but it is quite unlikely to have originated from accretion of many low-mass stellar systems (M$< 10^9$~\Msun). Accretion of satellites would most probably occur from various directions and significantly lower the overall stellar angular momentum, thus tending towards a slow(-er) rotating spheroid \citep{Vitvitska2002,Bournaud2007,Qu2010}, in striking contrast with what we observe within $4\,R_e$. Furthermore, the overall metallicity of the stars in the spheroid ([Z/H] $\sim -0.4$~dex) would be hard to explain if predominantly assembled via stellar accretion of low-mass systems (with an average [Fe/H] $\sim-1.5$~to~$-0.7$~dex for $10^{7-9}$~\Msun\ galaxies, \citealt{Kirby2013}). The more metal-rich cores of low-mass galaxies may match the required metallicity \citep{Koleva2011, Guerou2015}, but this would then imply an even larger number of accreted systems (to account for the present day spheroid mass), hence constraining further their orbital angular momentum distribution.

As observationally suggested, stellar mass growth of galaxies below $\sim$\,$5\times10^{10}$~\Msun\ should be dominated by in-situ star formation, and galaxies more massive than $\sim$ $10^{11}$~\Msun\ require significant merger events to climb up the region of high mass and large effective radii \citep{Cappellari2013XX}. This was suggested to be true at all redshifts \citep{Rodriguez-Gomez2015}, although there is clearly a large galaxy-to-galaxy variation. According to their work, accretion of small stellar systems should account for $\sim$10~--~30\% of the total stellar mass of a galaxy in the mass range of NGC\,3115 (see their figure~4), mostly detectable at $\sim$4~R$_e$ (see their figure 10). \cite{Arnold2011} and \cite{Pastorello2015} also suggested a transition radius of $4\,R_e$ for NGC\,3115, based on an observed decreasing rotation and different metallicity content of the globular clusters population at large radii. This would be consistent with the lack of strong evidence within the MUSE field of view ($\sim$4\,R$_e$) for ex-situ star formation for NGC\,3115, the galaxy being also near the mass transition ($10^{11}$~\Msun) mentioned earlier.  

In such a context, an early merger event with few (thus relatively massive) progenitors would make more sense to explain the origin of the spheroidal component of NGC\,3115 (within $\sim$4~R$_e$). \cite{Arnold2011} investigated this scenario and could not exclude that finely tuned simulations could reproduce such galaxy kinematics, in particular the surprising well aligned rotation between the stellar disc and spheroid components. The flattened spheroidal component of NGC\,3115 is rotating quite rapidly, as beautifully illustrated by the MUSE stellar kinematics. Such a high angular momentum may simply be the remnant of the orbital angular momentum associated with (only a few) merging large systems. Progenitors of $10^{10}$~\Msun\ and higher would also be consistent with the relatively high metallicity of the spheroid within 4~$R_e$ \citep{McDermid2015}. An early gas-rich merger event could thus explain the flattened spheroid, with the remnant gas reservoir filling in the rebuilding of a stellar disc structure \citep[see e.g.\,,][]{Athanassoula2016} as observed in the MUSE data.

Finally, even though we can not pinpoint the exact origin and evolution history of NGC\,3115, it is clear that most of its stellar mass formed at a redshift z>3, creating the roundish distribution of metal-poor stars we see nowadays (within 4~R$_e$), followed by a flattening of its spheroid (perhaps from a few merger events) by redshift z$\sim$2. Since then, NGC\,3115 evolved nearly secularly through dynamical processes (mostly associated with the discy potential). We thus speculated on the role of a potential bar which could have redistributed gaseous and stellar mass and driven the formation of rings and spirals.

%%%%%%%%%%%%%%%%%%%%%%%%%%%%%%%%%%%%%%%%%%%%
\section{Conclusion}
\label{sec:conclusion}

We presented MUSE commissioning data of NGC\,3115, the most nearby S0 galaxy. This data set was obtained to demonstrate the great potential of MUSE IFU instrument, mounted on the ESO VLT telescope, to cover extra-galactic sources at large radii (4~--~5\,R$_e$) in a reasonable amount of time (1h OB in the present case). We showed the mosaicing capacity of the instrument for diffuse objects, emphasising that the data quality has since been further improved, thanks to a revised observation strategy and an updated reduction pipeline. 

Through a detailed analysis of the stellar kinematics and population of NGC\,3115, we revealed the two dimensional structures of the different components of this galaxy: the spheroid, the intermediate-scale disc including spiral-like structures and rings, the nuclear disc, and briefly discussed the potential presence of a tumbling bar. We also presented a stellar population analysis, which for the first time, allowed us to provide a two-dimensional view of the star formation history, specifically disentangling the present day stellar population distribution in different bins of age and metallicity. 

Within 4\,R$_e$ of NGC\,3115’s spheroid component, although we find no direct evidence for an ex-situ origin as expected from simulations for this galaxy mass range ($\sim$ $1.1\times10^{11}$~\Msun), we suggest an early gas-rich merger event to explain the metallicity and the high rotation of the spheroid. The MUSE data also points towards the fact that NGC\,3115 has evolved nearly secularly for the past 10~Gyr (since z$\sim$2) allowing the long term existence of a thin fast rotating stellar disc with a gradually decreasing star formation rate.

This IFU data set emphasises the increasing need for new analysis tools to make the most of the overwhelming amount of information available. In particular, one further step forward would be to more intimately connect the two-dimensional (gaseous and stellar) kinematics and stellar population information, looking at the multi-dimensional data set via the LOSVDs, age and metallicity bins, which should provide an unprecedented view of the formation and mass assembly of galaxies. We believe that such a chemo-dynamical perspective will be key to eventually solve some of the galaxy formation puzzles.

\begin{acknowledgements}
    This work is based on public data released from the MUSE commissioning observations at the VLT Yepun (UT4) telescope under Programme ID 60.A-9100(A). The authors want to warmly thank Joel Vernet, Fernando Selman and all Paranal staff for their enthusiastic support of MUSE during the commissioning runs, Fran\c{c}oise Combes for the nice discussion, Johan Richard and Sandro Tacchella for useful comments as well as Michele Cappellari for providing us with mass profile data. Finally, we thank the anonymous referee for his/her constructive report.
\end{acknowledgements}

%-------------------------------------------------------------------
\bibliographystyle{aa}
\bibliography{MUSE,MUSE_extra}

\end{document}